\documentclass[10pt,twocolumn,letterpaper]{article}

\usepackage{wacv}
\usepackage[accsupp]{axessibility}  
\usepackage{times}
\usepackage{epsfig}

\usepackage[nocompress]{cite}
\usepackage{url}
\usepackage[switch]{lineno}

\usepackage{amsmath,amssymb,graphicx,booktabs,adjustbox}
\usepackage{microtype}
\usepackage{soul}
\usepackage{enumitem}
\usepackage{array}
\newcolumntype{H}{>{\setbox0=\hbox\bgroup}c<{\egroup}@{}}
\usepackage{comment}
\usepackage{pifont} 
\usepackage{multirow, multicol}
\usepackage{subcaption}
\usepackage{commath} 
\usepackage[table,xcdraw,dvipsnames]{xcolor}
\usepackage{authblk}

\newcommand{\cmark}{\ding{51}} 
\newcommand{\xmark}{\ding{55}} 
\newcommand{\NA}{---} 

\def\wacvPaperID{159} 

\wacvalgorithmstrack   

\wacvfinalcopy 

\usepackage[pagebackref=true,breaklinks=true,colorlinks,bookmarks=false]{hyperref}

\usepackage[capitalize]{cleveref}
\crefname{section}{Sec.}{Secs.}
\Crefname{section}{Section}{Sections}
\Crefname{table}{Table}{Tables}
\crefname{table}{Tab.}{Tabs.}

\begin{document}

\newcommand{\revised}[1]{\textcolor{black}{#1}}
\newcommand{\red}[1]{\textcolor{red}{#1}}
\newcommand{\blue}[1]{\textcolor{blue}{#1}}

\title{Single-Image HDR Reconstruction by Multi-Exposure Generation}

\author[1]{Phuoc-Hieu Le}
\author[1,2]{Quynh Le}
\author[1]{Rang Nguyen}
\author[1]{Binh-Son Hua}
\makeatletter
\renewcommand\Authsep{\quad}
\renewcommand\Authands{\quad}
\renewcommand\AB@affilsepx{\quad\protect\Affilfont} 
\makeatother
\affil[1]{VinAI Research}
\affil[2]{University of California San Diego}

\twocolumn[{%
    \renewcommand\twocolumn[1][]{#1}%
    \vspace{-5mm}
    \maketitle
    \vspace{-5mm}
    \thispagestyle{empty}
    \begin{center}
        \captionsetup{type=figure}
    	\footnotesize
        \begin{minipage}[c]{0.975\textwidth}
        \centering
            \includegraphics[width=\linewidth]{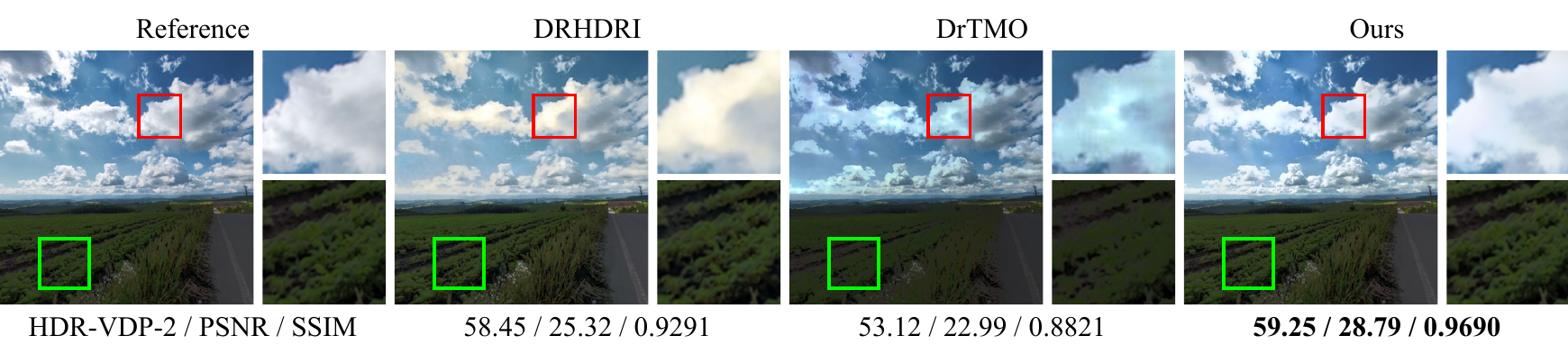}
        \end{minipage}
        \caption{The result from our weakly supervised single-image HDR reconstruction. Deep Recursive HDRI~\cite{drhdri} and DrTMO~\cite{drtmo} produce artifacts in saturated regions. Our method outperforms both previous methods with a more visually pleasing HDR.}
        \label{fig:teaser}
    \end{center}
}]

\begin{abstract}

\emph{High dynamic range}~(HDR) imaging is an indispensable technique in modern photography. 
Traditional methods focus on HDR reconstruction from multiple images, solving the core problems of image alignment, fusion, and tone mapping, yet having a perfect solution due to ghosting and other visual artifacts in the reconstruction. 
Recent attempts at single-image HDR reconstruction show a promising alternative: by learning to map pixel values to their irradiance using a neural network, one can bypass the align-and-merge pipeline completely yet still obtain a high-quality HDR image. 
In this work, we propose a weakly supervised learning method that inverts the physical image formation process for HDR reconstruction via learning to generate multiple exposures from a single image.
Our neural network can invert the camera response to reconstruct pixel irradiance before synthesizing multiple exposures and hallucinating details in under- and over-exposed regions from a single input image. 
To train the network, we propose a representation loss, a reconstruction loss, and a perceptual loss applied on pairs of under- and over-exposure images and thus do not require HDR images for training. 
Our experiments show that our proposed model can effectively reconstruct HDR images. Our qualitative and quantitative results show that our method achieves state-of-the-art performance on the DrTMO dataset.
Our code is available at \url{https://github.com/VinAIResearch/single_image_hdr}.

\end{abstract}

\section{Introduction} \label{introduction}

Cameras are optical instruments designed to mimic the \emph{human visual system} (HVS). 
They can capture the surrounding environment as close as what our eyes can observe in general conditions.
Unfortunately, an essential but challenging factor for the camera to reproduce compared to the human visual system is the dynamic range. 
Remarkably, dynamic ranges captured by a camera and our eyes are not the same. 
A consumer-grade camera can only capture images with relatively \emph{low dynamic ranges} (LDR images) while our eyes can perceive very \emph{high dynamic ranges} (HDR)~\cite{ferwerda1996model, photons2photos_2020}.
The images captured by such cameras often result in over-exposed regions with many saturated details. 

HDR imaging is a technique in modern photography to reproduce a higher dynamic range in a photograph so that more details of the bright and dark regions can be retained in the image. 
HDR images are a more faithful reproduction of a scene, being closer to HVS than traditional (low) dynamic range images. 
Beyond photography, HDR has found applications in image-based lighting, HDR display design, and computer vision downstream tasks.
As the need for HDR imaging becomes prevalent, techniques for HDR imaging reconstruction are requisite.

Unfortunately, acquiring HDR images is a challenging task. 
To reconstruct an HDR image, one typically needs special hardware like a camera system with HDR support; else, one has to capture multiple LDR images and reconstruct HDR computationally. 
The latter approach is more popular as its theories, and best practices have been well understood and widely implemented on consumer devices such as smartphones~\cite{hdrplus}. 
A common technique is to reconstruct HDR with multiple exposure images, where each exposure captures details in certain dynamic ranges. 
Notwithstanding such adoption, multi-exposure-based HDR suffers from visual artifacts due to object motion at capture. 
Computational photography methods on HDR reconstruction mainly deal with approaches that can mitigate these artifacts. 

With deep learning, reconstructing an HDR image from a single LDR image becomes a plausible solution to resolve visual artifacts caused by motion. 
In principle, we could generate a sequence of well-aligned images with different exposures, which can be fused by conventional methods~\cite{debevec, mertens} to generate an HDR image. Recent work~\cite{drtmo, lee2018deepchain, drhdri, lee2020learning} has made significant progress in this direction. However, these approaches are designed in a supervised learning manner that requires input images with corresponding ground truth. 
These approaches do not explicitly handle the missing-detail issue in saturated regions (as shown in \cref{fig:teaser}).
In this work, we propose a novel weakly supervised learning method that utilizes only low-dynamic range images from the same exposure stack for training HDR reconstruction. The basic idea is to learn to generate multiple exposure images from a single image by inverting the camera response and hallucinating missing details using neural networks. Our main contributions are: 
\begin{itemize}[leftmargin=*]
    \item \revised{A novel end-to-end trainable neural network that can generate an arbitrary number of different exposure images for HDR reconstruction. Our network is designed with weakly supervised learning that only uses multiple exposures for training, and thus can relax the requirement of obtaining ground truth HDR images;}
    \item An objective function that utilizes pixel irradiance for supervising the network using only multiple exposure images without the need for ground truth HDR image;
    \item Comprehensive quantitative and qualitative evaluations with results showing that the proposed framework is comparable to existing models.
\end{itemize}
We will release our implementation, evaluation code, and pre-trained model upon publication.

\section{Related Work} \label{related-work}

\noindent\textbf{High-quality HDR from Multiple Exposures.}
HDR reconstruction is a long-studied problem in computer vision. 
The typical approach to this problem is to reconstruct HDR from multiple exposures as suggested by {Debevec and Malik~\cite{debevec}} or Mertens \etal~\cite{mertens}. 
The aforementioned methods, however, often fail to reconstruct the desired HDR image properly, leading to artifacts, ghosting, and tearing in the final HDR image, especially when motion is introduced in the scene. 
For a long time, research in HDR reconstruction was focused on mitigating such artifacts~\cite{hdrplus, sen2012robust, serrano2016convolutional, hu2013hdr}. 

In the modern era of deep learning, Kalantari \etal~\cite{kalantari2017deep} proposed the first learning-based method for HDR image reconstruction for dynamic scenes, which performs image alignment and merging with a convolutional neural network (CNN).
Later work~\cite{prabhakar2019fast, prabhakar2020towards, pu2020robust} followed the previous pipeline but replaced the conventional optical flow in the alignment step with CNN.
Others opt for an end-to-end network~\cite{wu2018deep, yan2019attention, yan2020nonlocal} or a generative adversarial neural network (GAN)~\cite{niu2021hdr, li2021uphdr} to solve this problem.
While these methods, as mentioned earlier, can produce high-quality HDR images, eliminating ghosting artifacts is still a challenging problem in the multiple exposure pipeline. 

\begin{table}[t]
    \centering
    \resizebox{\columnwidth}{!}{%
        \begin{tabular}{c|lHHcc}
            \toprule
            Approach                  & Method               & Input & Output & \shortstack{Dataset\\ available} & Training         \\ \midrule
            \multirow{4}{*}{\shortstack{\textbf{Direct}\\Input: LDR\\Output: HDR}}   & Eilertsen \etal~\cite{hdrcnn} & LDR   & HDR    & No                               & Supervised       \\
                                      & Yang \etal~\cite{yang2018image}     & LDR   & HDR    & No                               & Supervised       \\
                                      & Moriwaki \etal~\cite{moriwaki2018hybrid}     & LDR   & HDR    & No                               & Supervised       \\
                                      & Marnerides \etal~\cite{expandnet} & LDR   & HDR    & No   & Supervised       \\
                                      & Khan \etal~\cite{FHDR} & LDR   & HDR & No                               & Supervised       \\
                                      & Liu \etal~\cite{singlehdr}      & LDR   & HDR    & Yes                              & Supervised       \\
                                      & Santos \etal~\cite{santos2020single}   & LDR   & HDR    & No                               & Supervised       \\ \midrule
            \multirow{4}{*}{\shortstack{\textbf{Indirect}\\Input: LDR\\Output: LDRs}} & Endo \etal~\cite{drtmo}      & LDR   & LDRs   & Yes                              & Exposure stack       \\
                                      & Lee \etal~\cite{lee2018deepchain}      & LDR   & LDRs   & No                               & Exposure stack       \\
                                      & Lee \etal~\cite{drhdri}      & LDR   & LDRs   & No                               & Exposure stack       \\
                                      & Lee \etal~\cite{lee2020learning}      & LDR   & LDRs   & No                               & Exposure stack       \\\cline{2-6} 
                                      & Ours                 & LDR   & LDRs   & Yes                              & Exposure pair  \\ \bottomrule
        \end{tabular}
    }
    \caption{Summary of single-image HDR reconstruction.}
    \label{tab:summary}
    \vspace{-0.2in}
\end{table}

\noindent\textbf{Single-Image HDR Reconstruction.}
Using a single image for HDR reconstruction is, therefore, beneficial in that the misalignment problem can be circumvented.
Eilertsen \etal~\cite{hdrcnn} proposed to use CNN to predict the missing information lost in saturated regions caused by sensor saturation. 
Differing from the previous work, Yang \etal~\cite{yang2018image} enriched details in LDR images by using a CNN that first recovered an HDR image with missing details. Then it learned a tone mapping function that mapped from HDR to the LDR domain with the retrieved details. 
Other ideas, such as employing a hybrid loss~\cite{moriwaki2018hybrid}, combining local and global features~\cite{expandnet} or using Feedback Network~\cite{FHDR}, have been proposed in an attempt to output more realistic results.
Recently, Liu \etal~\cite{singlehdr} decomposed the HDR imaging problem into three sub-tasks based on modeling the reversed, HDR-to-LDR image formation pipeline: dynamic range clipping, non-linear mapping, and quantization. 
Note that a similar idea~\cite{brooks2019unprocessing, zamir2020cycleisp} to reverse the camera pipeline has been applied in the denoising task.
Santos \etal~\cite{santos2020single} suggested a feature masking mechanism to guide the network to focus on the valid information in the well-exposed regions rather than saturated ones to avoid causing ambiguity during training the CNN. 
Furthermore, their work also suggested that pre-training with inpainting can help the network synthesize visually pleasing textures in the saturated areas.
The previous work typically extends the dynamic range of the input by \emph{directly} predicting details in the overexposed regions caused by sensor saturation. 
These methods require \emph{ground truth HDRs} available in the learning process to achieve that goal.

An \emph{indirect} way to reconstruct HDR from a single image is via the prediction of multiple exposure images. 
The final HDR photo is then reconstructed from the inferred bracketed LDR images. 
The benefit of this approach is that it allows more fine-grained control of the details by having the low- and high-exposure generation in separate processes. 
This idea was first explored by Endo \etal~\cite{drtmo}, where they used two neural networks to infer up- and down-exposure images from an LDR image with medium exposure. 
Similarly, Lee \etal~\cite{lee2018deepchain} later proposed a single model containing six sub-networks in a chaining structure to infer the bracketing images sequentially. With the input at a middle exposure value ($\text{EV}_0$), their model can infer EV ±1, ±2, ±3 effectively as the network goes deeper. 
As the number of synthesized bracketed images is fixed along with the exposure of each image, any attempts to overcome these limitations may require re-training the network. 
To overcome the mentioned issue, Lee \etal~\cite{drhdri} defined two neural networks representing the relationship between images with relative EVs. 
The proposed structure can scale well with the number of generated images without the need to re-train or to add more sub-networks.
Following that, Lee \etal~\cite{lee2020learning} improved the predecessor work by using two conditional GAN structures~\cite{mirza2014conditional} to generate a multi-exposure stack recursively. 
Although the mentioned frameworks can synthesize plausible multi-exposure stacks, it still has limitations since it has neither more granular control over the output exposure nor takes into account the image formation pipeline.
Our method is based on this indirect approach in that our network predicts multiple exposure images granularly by inverting the physical image formation pipeline.

\revised{
We summarize all methods for single-image HDR reconstruction in \cref{tab:summary}, where our method is based on weak supervision from multi-exposure images. 
Note that while the method of Ram \etal~\cite{deepfuse} is unsupervised, our method is different from that as we learn to generate multi-exposure stacks, whereas Ram \etal's method is used to fuse them.
}

\section{Our Approach} \label{our-method}

\subsection{Problem formulation}

\begin{figure}[t]
    \centering
    \includegraphics[width=\linewidth]{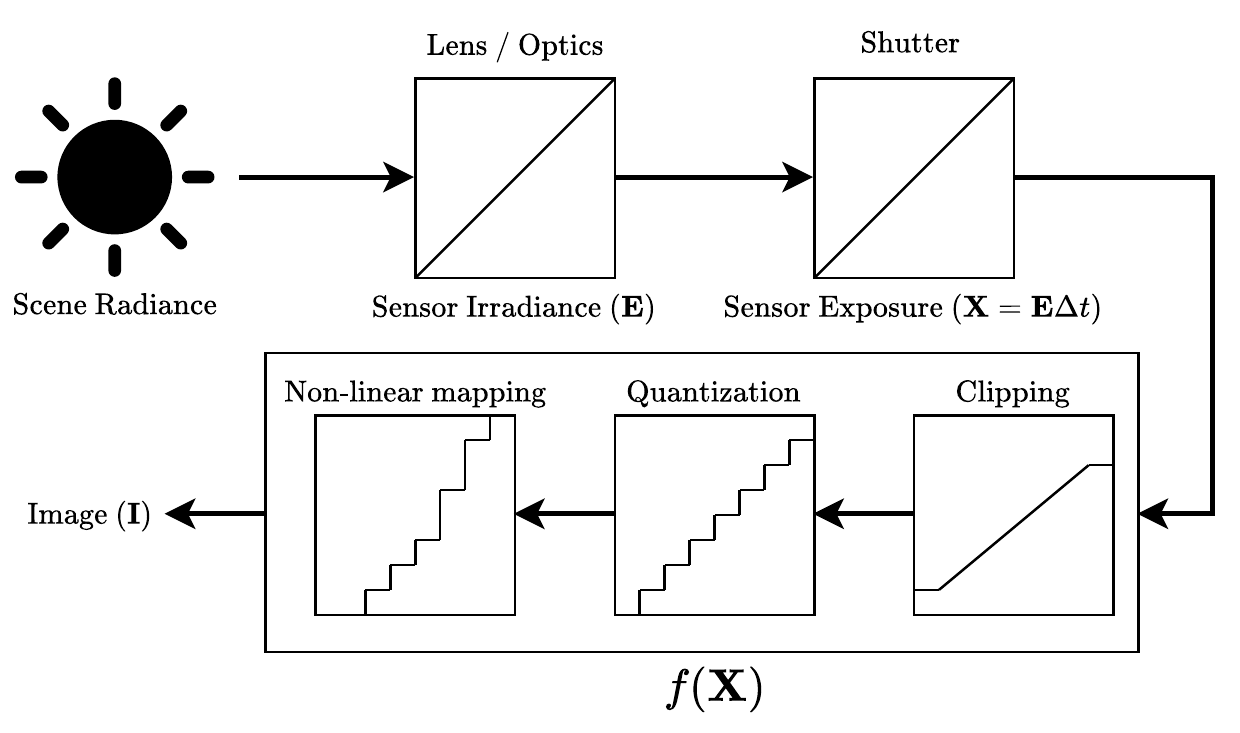}
    \caption{
        We model the traditional image formation pipeline by a non-linear function $f(X)$ where $X = E\Delta t$ is the sensor exposure formed by the product of sensor irradiance $E$ and exposure time $t$. 
        Our method learns to invert this function by using CNNs in an end-to-end fashion.
    }
    \label{fig:camera-pipeline}
    \vspace{-0.2in}
\end{figure}

\begin{figure*}[t]
    \centering
    \includegraphics[width=\linewidth]{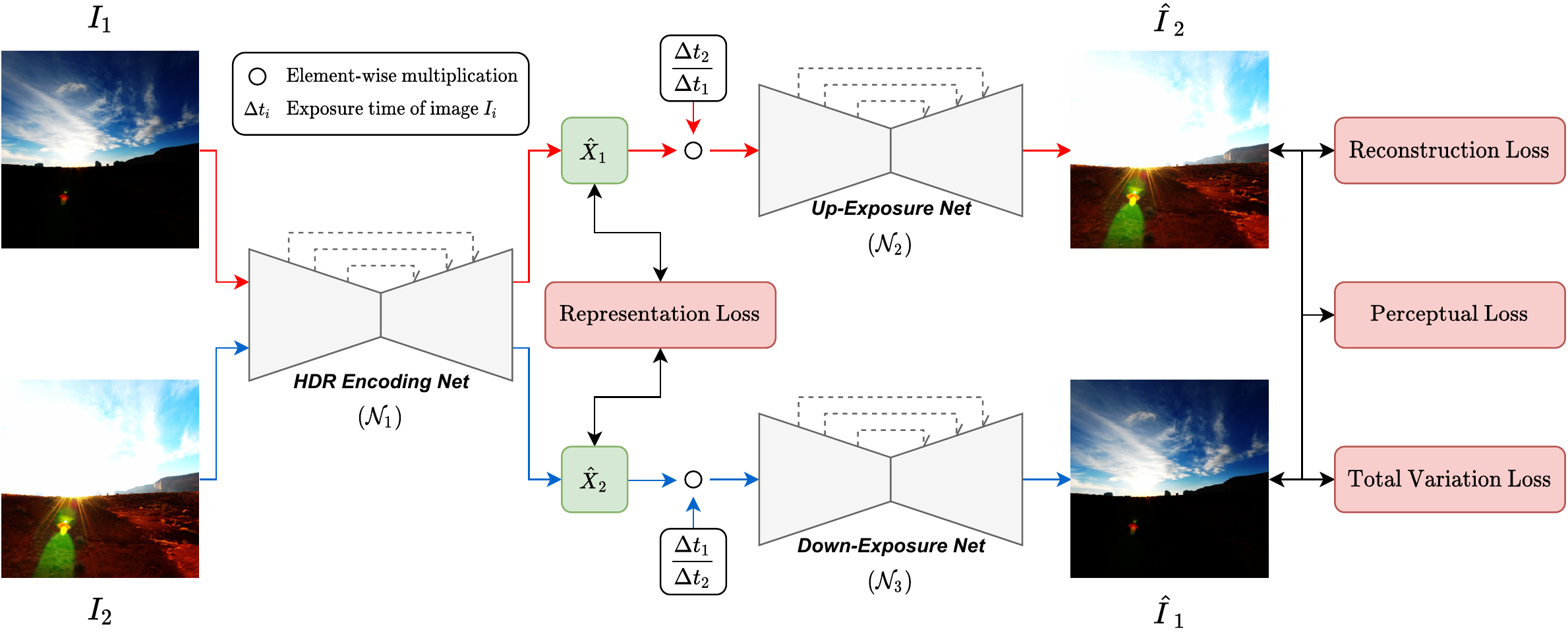}
    \caption{Training pipeline of our proposed framework. Given a pair of images in two different exposures, we predict latent invariant representation from the exposures by enforcing the exposure pair $(\hat{X}_1, \hat{X}_2)$ to have the same representation when scaled by a factor (network $N_1$). This representation can then be scaled and passed to Up/Down-Exposure Net ($N_2$ and $N_3$) to reconstruct different exposure images.}
    \label{fig:training-pipeline}
\end{figure*}

In this section, we propose our method for HDR image reconstruction. 
The basic idea is to let the network learn to generate multiple exposures from a single input image, and the HDR can be reconstructed from the generated exposures following the conventional HDR pipeline. 
Let us begin with our camera pipeline for image formation (\cref{fig:camera-pipeline}).
We generally model an image $I$ from the in-camera image processing pipeline as a function $f(X)$ that transforms the scene irradiance $E$ integrated over exposure time $\Delta t$. 
The light measurement $X = E\Delta t$ is linear and based on the physics model of light transport, and the function $f(X)$ is the composition of all non-linearities that occurred in the in-camera signal processing, such as camera response function, clipping, quantization, and other non-linear mappings.
In this model, the sensor irradiance $E$ captures the high dynamic range signals of the scene, and the image $I$ represents the low dynamic range (LDR) signal for display.
Note that we assume the noise in the process is insignificant. Thus we decide to exclude it from the pipeline.

To perform HDR reconstruction, our goal is to invert this image formation model to recover sensor irradiance $E$ from image $I$ so that $E = f^{-1}\left( I \right) / \Delta{t}$. 
This means we have to invert the non-linearities captured in $f(X)$. 
Unfortunately, this is an ill-posed and challenging problem because some steps in $f(X)$ are irreversible, e.g., clipping, while the camera response function varies per camera and is usually considered proprietary.
To address this problem, we opt for data-driven approaches, and we propose to use CNNs to learn the pipeline inversion. 

\subsection{Proposed network} \label{subsec:proposed-network}
Our HDR supervision is based on pairs of images taken with different exposures. We assume that the images are taken with the same camera in the same scene, so the images share the same underlying scene irradiance, which is also the same assumption as Grossberg and Nayar's method~\cite{grossberg2003determining}. Let the multiple exposures be $\left\{ I_i \right\}$ with 
$
    I_i = f\left( X_i \right) = f\left( E\Delta{t}_{i} \right), ~\forall i = 1, \dotsc, n
$
where $\Delta{t}_{i}$ is the exposure time of a corresponding image $I_i$ and sensor exposure $X_i = E \Delta t_i$ to supervise the neural network.
Particularly, for every pair of low- and high-exposure $\left( I_1, I_2 \right)$ from the same sensor irradiance $E$ and mapping function $f$ with corresponding exposure time $\left( \Delta{t}_1, \Delta{t}_2 \right)$ where $\Delta t_2 > \Delta t_1$ as input, we predict the up- and down-exposure $\hat{I}_2$ and $\hat{I}_1$ for $I_1$ and $I_2$, respectively. 
This task guides the network to generate $\hat{I}_1$ and $\hat{I}_2$ such that they match well the input exposure $I_1$ and $I_2$.
Mathematically, the relation between $I_1$ and $I_2$ can be written as:
\begin{align}
    X_i = E\Delta{t_i} = f^{-1}\left( I_i \right) 
    \label{eq:iTMO}
\end{align}
where $i \in \{1, 2\}$.
Given $X_i$, we can then scale it accordingly and generate a different exposure image $I_j$:
\begin{align}
    I_j = f(X_j) &= f(E\Delta{t_j})
    = f\left(X_i\frac{\Delta{t_j}}{\Delta{t_i}}\right)
    \label{eq:transform-X}
\end{align}
where $j \in \{2, 1\}$.
\Cref{fig:training-pipeline} shows the overall structure of our proposed method.

\noindent\textbf{Network overview.}
Our proposed network consists of two stages. The first stage is backward mapping, where we use our \emph{HDR Encoding Net} $\left( \mathcal{N}_1 \right)$ to transform input image $I_i$ into $X_i$, a suitable representation for the image's sensor exposure at exposure time $\Delta{t}_i$ in latent space. An appropriate factor can scale the representation $X_i$, as shown in \cref{eq:transform-X} to get the sensor exposure at a shorter or longer exposure time $\Delta{t}_{i \pm 1}$. 

The next stage is forward mapping, where we want to map the scene irradiance $X_i$ into a pixel value after altering its exposure time $\Delta{t}_i$ to generate a new image with a different exposure. 
In this stage, it requires to hallucinate details in saturated regions. 
Due to the different nature of under- and over-exposed images, we propose to use two sub-networks \emph{Up-Exposure Net} $\left( \mathcal{N}_2 \right)$ and \emph{Down-Exposure Net} $\left( \mathcal{N}_3 \right)$ respectively to hallucinate and generate the high- and low-exposure image with respect to the input image following the \cref{eq:transform-X}. 

\noindent\textbf{Masked Regions.}
In our model, we use masks to eliminate the over- and under-exposed regions in the input image, as Santos \etal~\cite{santos2020single} suggested in their approach. 
We use masks in both the training and inference phase.
Please refer to the supplementary for more details and illustration about the mask generation process.

\subsection{Loss function} \label{sec:loss-function}
The proposed network can be trained in an end-to-end fashion. 
In our training, we sample a pair of low- and high-exposure $(I_1, I_2)$ as input, respectively. 
By applying the $\mathcal{N}_1$ network to recover the latent sensor irradiance $( \hat{X}_1, \hat{X}_2 )$, we then scale them by a factor $\frac{\Delta{t_i}}{\Delta{t_j}}$ like in \cref{eq:transform-X} before feeding the outputs to the $\mathcal{N}_2$ and $\mathcal{N}_3$ to get the predicted $( \hat{I}_2, \hat{I}_1 )$ exposure. 

The choice of appropriate loss functions is critical for the learning process of one's model. 
In order to synthesize realistic bracketed images, apart from typically used loss functions, we also introduce prior knowledge of the camera formation pipeline to our proposed network. 
Specifically, we train our network by optimizing the combination of HDR Representation loss $\mathcal{L}_h$, reconstruction loss $\mathcal{L}_r$, VGG perceptual loss $\mathcal{L}_p$ and total variation (TV) loss $\mathcal{L}_{tv}$. 
Mathematically, the final combined loss function $\mathcal{L}$ is:
\begin{align}
    \mathcal{L} = \lambda_{h}\mathcal{L}_{h} + \lambda_{r}\mathcal{L}_{r} + \lambda_{p}\mathcal{L}_{p} + \lambda_{tv}\mathcal{L}_{tv}
\end{align}

\noindent\textbf{HDR Representation Loss.}
With the setup for our training as shown in \cref{fig:training-pipeline} and the image formation pipeline in \cref{fig:camera-pipeline}, we know that $\left( I_1, I_2 \right)$ comes from the same scene irradiance $E$ with an identical CRF $f$. Mathematically, we have:
\begin{align}
    \begin{cases}
        I_1 &= f\left( X_1 \right) = f\left( E\Delta{t}_{1} \right) \\
        I_2 &= f\left( X_2 \right) = f\left (E\Delta{t}_{2} \right)
    \end{cases}
\end{align}
We could infer that, if the inverse CRF $f^{-1}$ is known, $I_1$ and $I_2$ only differ by scalars $\Delta{t}_1$ and $\Delta{t}_2$ respectively.  
Our method relies on this prior knowledge to constrain $\mathcal{N}_1$. 
We introduce the \emph{transformation loss} when transforming from $\hat{X_1}$ to $\hat{X_2}$, following the \cref{eq:transform-X}, as:
\begin{align}
    \mathcal{L}_t\left( \hat{X}_1, \hat{X}_2 \right) =
    \norm{
        \log\left( \hat{X}_1 \odot \dfrac{\Delta{t}_2}{\Delta{t}_1} + \epsilon \right) - \log\left( \hat{X}_2 + \epsilon \right)
    }_{1}
\end{align}
where $\Delta{t}_i$ is the exposure time of the corresponding $\hat{X}_i$ and $\epsilon$ is a small constant to prevent numerical error. 
We take the logarithm of both encoded sensor irradiance before computing loss as computing the loss in the log domain reduces the influence of these substantial errors and encourages the network to restore more details in other regions. 
Then the \emph{HDR Representation loss} is defined as:
\begin{align}
\label{loss:hdr-representation}
    \mathcal{L}_h = \mathcal{L}_t\left( \hat{X}_1, \hat{X}_2 \right) + \mathcal{L}_t\left( \hat{X}_2, \hat{X}_1 \right)
\end{align}
with $\hat{X}_1$, $\hat{X}_2$ is the output from $\mathcal{N}_1$ given the input is $I_1$, $I_2$ respectively. 
The $\mathcal{L}_h$ loss can be seen as a relaxed version instead of directly forcing $\hat{X}_1$, $\hat{X}_2$, e.g., $\norm{\frac{\hat{X}_1}{\Delta{t}_1} - \frac{\hat{X}_2}{\Delta{t}_2}}_{1}$. 
If our network can learn to predict $\hat{X}_i = X_i$, the loss function would reach its minimum. 
The reason for using this loss function is that directly constraining $\hat{X}_i$ by multiplying or dividing could lead to exploding or vanishing gradient since $\Delta{t}$ usually lies within $\left[ 0, 1 \right]$.
This would make the training very unstable. 
Therefore, the loss function in \cref{loss:hdr-representation} that we have derived is more suitable for training.

\noindent\textbf{Reconstruction Loss.}
For supervising up and down-exposure networks ($\mathcal{N}_2$ and $\mathcal{N}_3$), this can be seen as an image-to-image translation task in which the typical losses used are pixel-wise $\ell_1$-norm and $\ell_2$-norm. Previous work~\cite{yan2019attention, zhao2016loss} has shown that $\ell_1$ is more effective for preserving details, thus we employ our \emph{reconstruction loss} as:
\begin{align}
    \mathcal{L}_r = 
    \norm{\hat{I}_1 - I_1}_{1}
    + 
    \norm{\hat{I}_2 - I_2}_{1}
\end{align}
where $\hat{I}_2$, $\hat{I}_1$ is the output from $\mathcal{N}_2$, $\mathcal{N}_3$ given the input image is $I_1$ and $I_2$ respectively.

\noindent\textbf{Perceptual Loss.}
The \emph{perceptual loss} is used to evaluate how well the extracted features from prediction are matched with the extracted features from the ground truth. This can help alleviate artifacts and create more realistic details in inferred images. Following setup in~\cite{santos2020single, singlehdr} we define our \emph{perceptual loss} as follows:
\begin{align}
    \mathcal{L}_p = 
    \displaystyle{
        \sum_{l}{\norm{\phi_{l}\left( \hat{I}_1 \right) - \phi_{l}\left( I_1 \right)}_1}
        +
        \sum_{l}{\norm{\phi_{l}\left( \hat{I}_2 \right) - \phi_{l}\left( I_2 \right)}_1}
    }
\end{align}
where $\phi_{l}\left( \cdot \right)$ is the extracted features at the $l^\text{th}$ layer of VGG network. In this work, we use VGG-19 network~\cite{vgg} and feature vectors are extracted from \emph{pool1, pool2, pool3} layers.

\noindent\textbf{Total Variation Loss}
To avoid overfitting when training and improve spatial smoothness of inferred images, we also minimize the \emph{total variation loss}. The total variation~\cite{totalvariation} for an image $y$ can be expressed as:
\begin{align}
    \mathbf{V}\left( y \right) = \displaystyle{
        \sum_{i, j} 
        \left(
                \norm{y_{i+1, j} - y_{i, j}}_1
                +
                \norm{y_{i, j+1} - y_{i, j}}_1
        \right)
    }
\end{align}
where $i, j$ are the corresponding pixel coordinates of that image. Given the above definition, our \emph{total variation loss} is calculated on inferred images as:
\begin{align}
    \mathcal{L}_{tv} = \mathbf{V}(\hat{I}_1) + \mathbf{V}(\hat{I}_2)
\end{align}

\subsection{Inference process}
To perform inference, we let the network take \emph{a single LDR image} as input and produce multiple up- and down-exposures of the input image to synthesize an image bracket. 
Specifically, different exposures are synthesized by scaling the latent scene irradiance of the input with an exposure ratio as discussed in \cref{subsec:proposed-network}.
We can then aggregate the images in the bracket to form an HDR image by following the conventional HDR imaging pipeline~\cite{debevec, mertens}. 
In this work, we use Photomatix~\cite{Photomatix} to generate the HDR image. Tone mapping is then followed to display the HDR image. 
More details of the inference process are provided in the supplementary.

\section{Experimental Results} \label{experimental-results}

\subsection{Implementation details}

\begin{figure*}[t!]
    \centering
    \includegraphics[width=\linewidth]{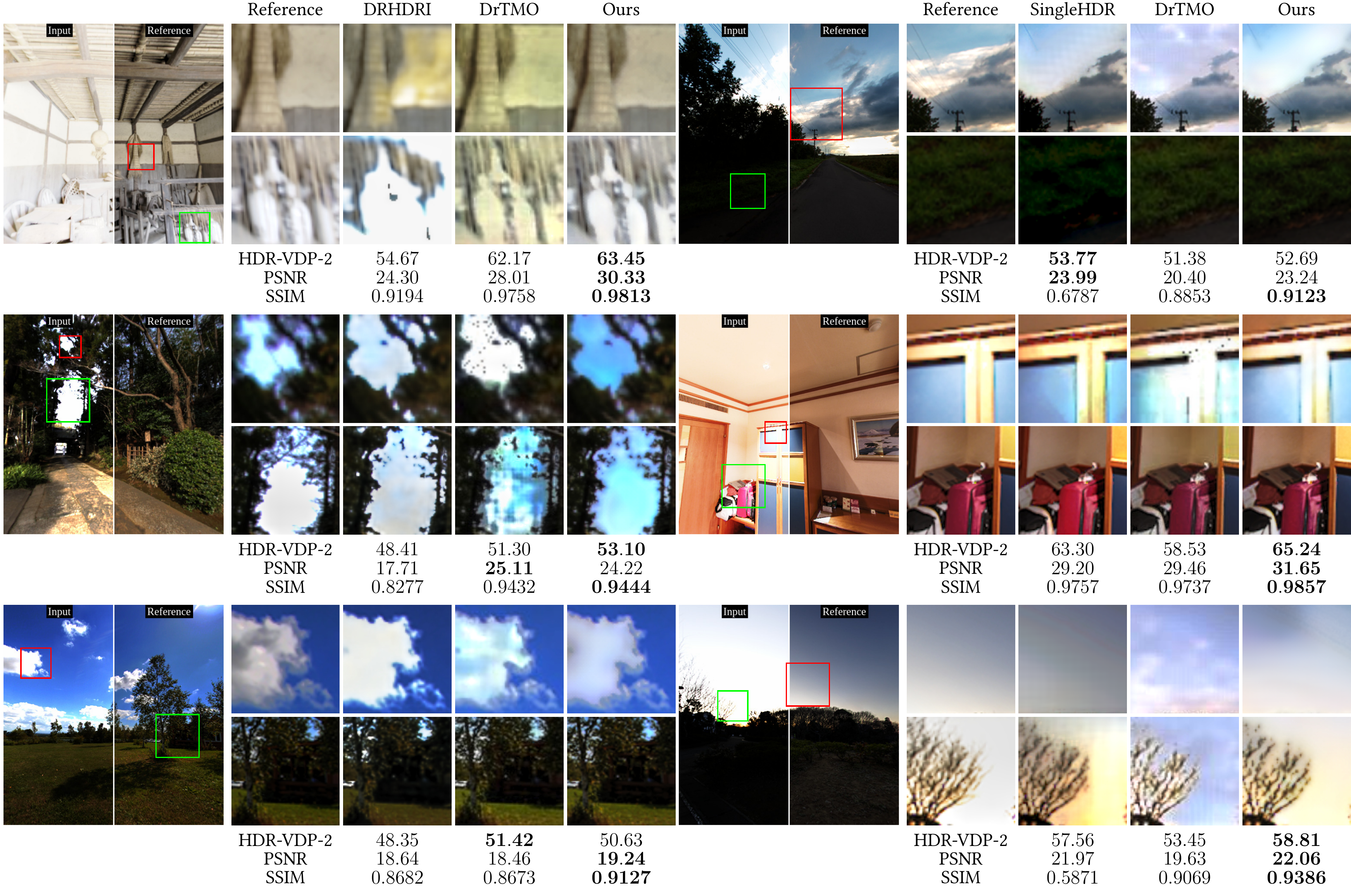}
    \caption[Tone-mapped HDR images comparison between different methods]{Tone-mapped HDR images comparison between different methods. DrTMO~\cite{drtmo} and Deep Recursive HDRI~\cite{drhdri} produce artifacts in extremely high dynamic range regions, SingleHDR~\cite{singlehdr} appears to have checkboard artifacts, while our method can recover details in these regions pleasingly.}
    \label{fig:compare-tone-mapped}
\end{figure*}

\noindent\textbf{Dataset.}
We used the dataset synthesized by Endo \etal~\cite{drtmo} for training and testing. 
The dataset was created by applying five representative CRFs from Grossberg and Nayar's Database of Response Functions (DoRF)~\cite{grossberg2003space} on 1,043 collected HDR images with nine exposure values. 
The process results in a total of 46,935 LDR images for training and 6,210 images for testing. 
Each image has a size of $512\times512$. 
Images in the dataset cover a wide variety of scenarios such as indoor, outdoor, night, and day scenes. 
Since our training pipeline receives only pairs of images, we randomly sample two images from each scene in the training set and use them as input to train our model.
We do not evaluate on other datasets such as HDR-S\textsc{ynth}, HDR-R\textsc{eal} from Liu \etal~\cite{singlehdr}, and RAISE~\cite{dang2015raise} because they do not include image pairs with known exposures that are required for our training.
Specifically, we investigated the HDR-S\textsc{ynth}, HDR-R\textsc{eal} data and found that their multiple exposure images are not well organized. The exposures do not match the index on file names, and the images do not include any EXIF tags that can be used for recovering the exposure information.

\begin{table}
    \centering
    \resizebox{\columnwidth}{!}{%
        \begin{tabular}{lcccc}
            \toprule
        
            Method & PSNR $\left( \uparrow \right)$ & SSIM $\left( \uparrow \right)$ & LPIPS $\left( \downarrow \right)$ & HDR-VDP-2 $\left( \uparrow \right)$  \\ \midrule

            Lee \etal~\cite{drhdri}     & 19.56         & 0.7920        & 0.2096        & 53.86 $\pm$ 4.46 \\
            Endo \etal~\cite{drtmo}     & \blue{21.60}  & \blue{0.8493} & 0.1592        & 54.56 $\pm$ 4.29 \\
            Liu \etal~\cite{singlehdr}  & 19.77         & 0.7832        & 0.2001        & 52.77 $\pm$ 5.40 \\
            Liu \etal~\cite{singlehdr}* & 21.58         & 0.8333        & \blue{0.1427} & \red{56.42 $\pm$ 4.50} \\
            Ours                        & \red{23.74}   & \red{0.8916}  & \red{0.1231}  & \blue{55.69 $\pm$ 5.01} \\
            
            \bottomrule
        \end{tabular}
    }
    \caption[Quantitative comparisons on HDR images]{Quantitative comparisons on HDR images. \red{Red} and \blue{blue} text indicates the best and second-best respectively. * indicates that the model has been pretrained on HDR-S\textsc{ynth}~\cite{singlehdr}.}
    \label{tab:compare-hdr}
    \vspace{-0.2in}
\end{table}

\noindent\textbf{Training details.}
Our model is trained using Adam optimizer~\cite{adam} with batch size and learning rate as $64$ and $1\times10^{-4}$, respectively.
We decrease the learning rate by a factor of $0.5$ every time the loss reaches a plateau. 
For each image in the input pair, we randomly crop a patch of $256\times256$ from it. 
The cropped image also gets randomly rotated, shifted, scaled, and flipped horizontally and vertically for augmentation, which enriches the input data and prevents the model from overfitting. 
We implement our model using PyTorch~\cite{pytorch} and train it on 2$\times$NVIDIA Tesla A100 GPU with approximate $200.000$ steps for our model to converge. 
The training phase took about three days to complete.

\subsection{Evaluation results}

\begin{table*}
    \centering
    \small
    \begin{tabular}{lccccccc}
        \toprule
        
        \multirow{2}{*}{Method} & \multicolumn{3}{c}{Reinhard \etal~\cite{reinhard}} && \multicolumn{3}{c}{Photomatix~\cite{Photomatix}} \\ \cmidrule{2-4} \cmidrule{6-8}
        & PSNR $\left( \uparrow \right)$ & SSIM $\left( \uparrow \right)$ & LPIPS $\left( \downarrow \right)$ && PSNR $\left( \uparrow \right)$ & SSIM $\left( \uparrow \right)$ & LPIPS $\left( \downarrow \right)$  \\ \midrule
        Lee \etal~\cite{drhdri}     & 22.68         & 0.9017        & 0.1438        && 21.42        & 0.8730        & 0.1785 \\
        Endo \etal~\cite{drtmo}     & \blue{27.19}  & \blue{0.9419} & 0.1289        && 22.20        & \blue{0.8944} & 0.1642 \\
        Liu \etal~\cite{singlehdr}  & 23.22         & 0.8954        & 0.1400        && 19.90        & 0.8548        & 0.1651 \\
        Liu \etal~\cite{singlehdr}* & 26.19         & 0.9135        & \blue{0.0969} && \blue{23.89} & 0.8856        & \blue{0.1236} \\
        Ours                        & \red{29.68}   & \red{0.9586}  & \red{0.0617}  && \red{25.22}  & \red{0.9370}  & \red{0.0933} \\
        
        \bottomrule
    \end{tabular}
    \caption[Quantitative comparisons on LDR images]{Quantitative comparisons on tone-mapped images with existing methods. \red{Red} and \blue{blue} text indicates the best and second-best respectively. * represents that the model has been pretrained on HDR-S\textsc{ynth}~\cite{singlehdr}. The proposed method outperforms all the others.}
    \label{tab:compare-ldr}
\end{table*}

\begin{table*}
    \centering
    \resizebox{\textwidth}{!}{%
        \begin{tabular}{cc|cccc|ccc|ccc}
            \toprule
              \multicolumn{2}{ c|}{Component} &
              \multicolumn{4}{ c|}{HDR} &
              \multicolumn{3}{ c|}{Reinhard Tonemap} &
              \multicolumn{3}{ c }{Photomatix Tonemap} \\ \midrule
              \textbf{Two Exposure Nets} &
              \textbf{HDR Loss} &
              PSNR (↑) &
              SSIM (↑) &
              LPIPS (↓) &
              HDR-VDP-2 (↑) &
              PSNR (↑) &
              SSIM (↑) &
              LPIPS (↓) &
              PSNR (↑) &
              SSIM (↑) &
              LPIPS (↓) \\ \midrule

              \xmark &
              \xmark &
              17.51 &
              0.5068 &
              0.3169 &
              52.41 $\pm$ 5.54 &
              15.99 &
              0.7764 &
              0.2421 &
              19.04 &
              0.8009 &
              0.2585 \\  

              \xmark &
              \cmark &
              14.48 &
              0.4068 &
              0.3464 &
              50.98 $\pm$ 5.43 &
              14.41 &
              0.7571 &
              0.2107 &
              17.10 &
              0.7804 &
              0.2359 \\  
             
              \cmark &
              \xmark &
              23.05 &
              0.8868 &
              \textbf{0.1192} &
              \textbf{56.68 $\pm$ 4.81} &
              29.29 &
              \textbf{0.9608} &
              \textbf{0.0580} &
              25.19 &
              \textbf{0.9396} &
              \textbf{0.0873} \\ 
            
              \cmark &
              \cmark &
              \textbf{23.74} &
              \textbf{0.8916} &
              0.1231 &
              55.69 $\pm$ 5.01 &
              \textbf{29.68} &
              0.9586 &
              0.0617 &
              \textbf{25.22} &
              0.9370 &
              0.0933 \\ \bottomrule
        \end{tabular}
    }
    \caption{Ablation study. Using separate exposure networks or HDR representation loss alone leads to degraded performance.}
    \label{tab:ablation}
\end{table*}

\noindent\textbf{Evaluation protocol.}
To demonstrate our model's ability to generate realistic images, we conduct experiments to compare our method against Endo \etal (DrTMO)~\cite{drtmo}, Lee \etal~\cite{drhdri} (DRHDRI), and Liu \etal~\cite{singlehdr}. 
We also considered Deep Chain HDRI~\cite{lee2018deepchain} but could not compare due to the lack of publicly available implementation. 
The model proposed by Lee \etal~\cite{drhdri} and their evaluation protocol only use five images with EV ranging from -2 to +2 given the input image has the EV of 0, each image is different by 1 EV, to reconstruct the HDR image. 
Thus, we decided to use the same setup as them. 
For Endo \etal~\cite{drtmo}, the model produces a total of 16 images, with eight images for up and the rest for down-exposed images. 
We decided to select only five images with the EV difference in the range of -2 to +2 from these for a fair evaluation between different models.
For Liu \etal~\cite{singlehdr}, we include two versions: with and without pretraining on HDR-S\textsc{ynth} dataset~\cite{singlehdr}. 
Note that Liu \etal's method requires ground truth HDR for training.

\noindent\textbf{Comparisons on HDR images.}
We use Photomatix~\cite{Photomatix} to recover the final HDR image from the predicted exposures.
The peak signal-to-noise ratio (PSNR), structure similarity (SSIM), LPIPS~\cite{lpips}, and HDR-VDP-2~\cite{hdrvdp} metrics are used to evaluate our reconstructed HDR from bracketed images. 
The result is shown in \cref{tab:compare-hdr}. 
Our model outperforms all previous works in PSNR, SSIM, and LPIPS.
Compared to Liu \etal~\cite{singlehdr}, our model outperforms in the HDR-VDP-2 metric if their model is not pretrained on HDR-S\textsc{ynth}. Their pretrained version is slightly better in HDR-VDP-2 metric due to the use of extra training data.

\noindent\textbf{Comparisons on tone-mapped images.}
The tone mapping operator (TMO) that we use to map the reconstructed HDR image into displayable LDR one is Reinhard~\etal~'s method~\cite{reinhard}, a popular global TMO that models the human visual system. We also consider the TMO from Photomatix~\cite{Photomatix} to validate the consistency in quantitative results between different methods. 
\Cref{tab:compare-ldr} shows that our proposed method outperforms all of the others in terms of all metrics - PSNR, SSIM, and LPIPS.

\Cref{fig:compare-tone-mapped} shows our tone-mapped HDR images along with others. DrTMO~\cite{drtmo} and Deep Recursive HDRI~\cite{drhdri} produce artifacts in extremely high dynamic range regions. While SingleHDR~\cite{singlehdr} could handle these regions, the results appear to have checkboard artifacts. Our method can recover these regions with reasonable details and without artifacts. 
More qualitative results can be found in the supplementary material.

\Cref{fig:compare-stack-images-1} compares four different exposure values synthesized by Deep Recursive HDRI~\cite{drhdri}, DrTMO~\cite{drtmo}, and ours, along with the corresponding ground truth. In the lowest EV, our model can synthesize details in the blue sky that is most plausible and near the ground truth image than the other two methods as DrTMO~\cite{drtmo} seems to suffer from artifacts when trying to fill in the details. While in the up-exposure scenario, all considered methods perform reasonably well, but the color and contrast of Deep Recursive HDRI~\cite{drhdri} do not match the ground truth images. Their network does not seem to model the CRF well enough to preserve the specific non-linear mapping in the CRF.

\begin{figure*}
    \centering
    \includegraphics[width=0.75\linewidth]{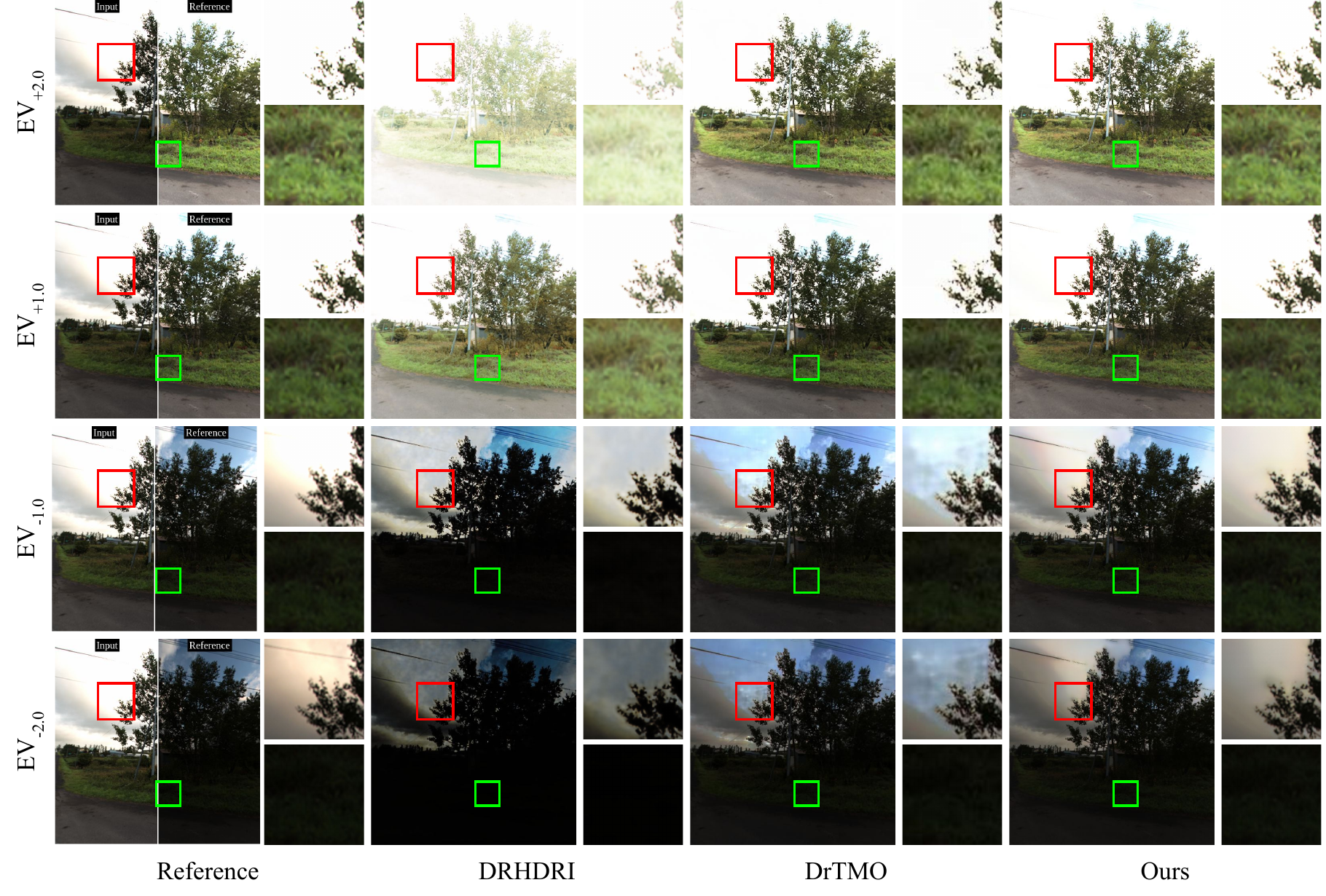}
    \caption{
    Comparison of bracketed images generated by our model, DrTMO~\cite{drtmo}, Deep Recursive HDRI~\cite{drhdri}, and the reference. 
    The overall structure of images is well reconstructed as well as perceptually similar to ground truth images for our method. 
    }
    \label{fig:compare-stack-images-1}
\end{figure*}

\noindent\textbf{Ablation study.}
We provided an ablation study in \cref{tab:ablation} to highlight the effectiveness of our proposed components.
Similar to the previous works~\cite{drtmo, lee2020learning, drhdri, lee2020learning}, we empirically found that using a single network to learn up-/down-exposure is ineffective.
This study also confirms the benefit of our proposed HDR loss for the PSNR metric.

\begin{table}[t!]
    \centering
    \resizebox{\columnwidth}{!}{%
        \begin{tabular}{c|ccc}
            \toprule

            Masked over-/under-exposure                 & PSNR $\left( \uparrow \right)$    & SSIM $\left( \uparrow \right)$    & LPIPS $\left( \downarrow \right)$ \\ \midrule

            \xmark                                          & 16.62                             & 0.8423                            & 0.1681                            \\
            \cmark                                          & 16.67                             & 0.8581                            & 0.1391                            \\

            \bottomrule
            \end{tabular}
    }
    \caption{Analysis on HDR Encoding Net ($\mathcal{N}_1$).}
    \label{tab:N1-analysis}
    \vspace{-0.2in}
\end{table}

\begin{figure}
    \centering
    \includegraphics[width=\linewidth]{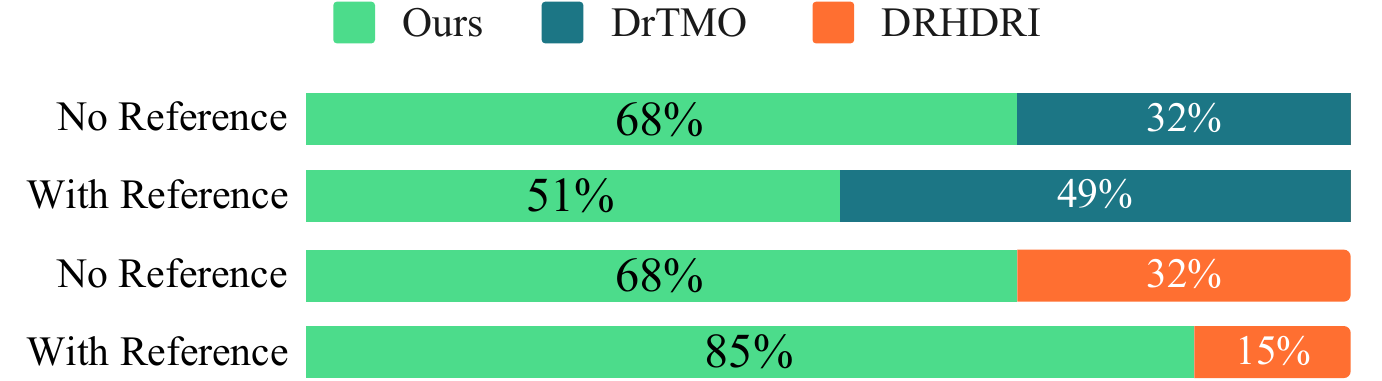}
    \caption{Results of user study - Ours vs. the Others. The proposed method is preferred to other methods in both tests.}
    \label{fig:user-study}
    \vspace{-0.2in}
\end{figure}

\noindent\textbf{Network structure analysis.}
To understand our proposed network, we provide an analysis of the effect of our HDR Encoding Net (network $\mathcal{N}_1$ in \cref{fig:training-pipeline}). 
We take an LDR image $I$ as input to $\mathcal{N}_1$. Each LDR image in the dataset is associated with a ground truth camera response function $g$, which we use to apply to the output of $\mathcal{N}_1$. We evaluate the similarity of $I$ and $g(\mathcal{N}_1(I))$ over all images in the test set using PSNR, SSIM, and LPIPS. 
The results are shown in \cref{tab:N1-analysis}. 
It can be seen that $\mathcal{N}_1$ recovers the scene radiance to a certain extent, reflected by a mid-range PSNR, SSIM, and LPIPS. Visualizations of the output of $\mathcal{N}_1$ can be found in the supplementary material. 

\subsection{User study}
We conduct a user study on 40 samples to evaluate human preference on the qualitative results. We randomly pick 28 scenes from the total of 33 scenes and show each in pairs~\cite{lai2016comparative, rubinstein2010comparative}. The participants are asked to pick a better image in each pair. First, in the test without reference, we show only two images, one is ours, and one is from the other method to evaluate \emph{visual quality} of the two HDR reconstructions. Then we do a reference test, in which we add an input LDR and a ground truth HDR image to each question. This is to evaluate the \emph{faithfulness} of each method. We report the detailed comparison in \cref{fig:user-study}, which shows that we are preferred in both tests compared to all other methods by approximately 70\% of the users.

\section{Conclusion} \label{conclusion}

We proposed a method for predicting multiple exposures from a single input image, which is applied for HDR image reconstruction. 
Our method contributes toward making data-driven HDR reconstruction more accessible without the need for ground-truth HDR images for training.
Our method can generate an arbitrary number of different exposure images, enabling more granular control in the HDR reconstruction. 
We achieved state-of-the-art performance on DrTMO dataset~\cite{drtmo}. 

A limitation of our method is that our reconstruction might have visual artifacts or missing details, which we hypothesize this is due to the diversity of the DrTMO dataset as it includes both natural outdoor and man-made indoor scenes, making the hallucination extremely challenging to learn. 
Future research might integrate generative modeling to improve the image quality or condition the reconstruction on example photographs. 

\noindent\textbf{Acknowledgment.} This work is done when Quynh Le was a resident of the AI Residency program at VinAI Research.

{\small
\bibliographystyle{ieee_fullname}
\bibliography{egbib}
}

\clearpage
\begin{center}\textbf{\Large Supplementary Material}\end{center}
\appendix

\begin{abstract}
    In this supplemental document, we show additional comparisons on HDR merging methods (\cref{sec:hdr_merging_comparisons}), on multiple-exposure images predicted by the network (\cref{sec:more_comparisons}) and then provide details about the proposed network, including network architecture (\cref{sec:network_architecture}) and the inference process (\cref{sec:inference-process}). 
\end{abstract}

\begin{table*}
    \centering
    \resizebox{\textwidth}{!}{%
        \begin{tabular}{lcccccccccccccc}
            \toprule
              \multirow{2}{*}{\stackbox{HDR Reconstruction\\ Method}} &
              \multicolumn{4}{c}{Photomatix~\cite{Photomatix}} &&
              \multicolumn{4}{c}{Debevec and Malik~\cite{debevec}} &&
              \multicolumn{4}{c}{Robertson~\etal~\cite{robertson1999merge}} \\ \cmidrule{2-5} \cmidrule{7-10} \cmidrule{12-15} &
              PSNR (↑) &
              SSIM (↑) &
              LPIPS (↓) &
              HDR-VDP-2 (↑) &&
              PSNR (↑) &
              SSIM (↑) &
              LPIPS (↓) &
              HDR-VDP-2 (↑) &&
              PSNR (↑) &
              SSIM (↑) &
              LPIPS (↓) &
              HDR-VDP-2 (↑) \\ \midrule

              Lee \etal~\cite{drhdri} &
              19.56 &
              0.7920 &
              0.2096 &
              53.86 $\pm$ 4.46 &&
              21.07 &
              0.9017 &
              0.1362 &
              39.21 $\pm$ 2.94 &&
              \textbf{19.20} &
              \textbf{0.8363} &
              0.2012 &
              39.03 $\pm$ 2.79 \\

              Endo \etal~\cite{drtmo} &
              21.60 &
              0.8493 &
              0.1592 &
              54.56 $\pm$ 4.29 &&
              24.27 &
              0.9243 &
              0.1404 &
              39.32 $\pm$ 2.93 &&
              15.27 &
              0.7342 &
              0.2664 &
              39.09 $\pm$ 2.84 \\

              Ours &
              \textbf{23.74} &
              \textbf{0.8916} &
              \textbf{0.1231} &
              \textbf{55.69 $\pm$ 5.01} &&
              \textbf{25.67} &
              \textbf{0.9434} &
              \textbf{0.0802} &
              \textbf{39.42 $\pm$ 2.95} &&
              16.34 &
              0.8035 &
              \textbf{0.1722} &
              \textbf{39.30 $\pm$ 2.90} \\ \bottomrule
        \end{tabular}
    }
    \caption[Quantitative comparisons on HDR images]{Quantitative comparisons on HDR images using different HDR merging methods.}
    \label{tab:compare-hdr-addition}
\end{table*}

\begin{table*}[!ht]
    \centering
    \small
    \begin{tabular}{lccccccc}
        \toprule
        
        \multirow{2}{*}{Tone-mapping Method} & \multicolumn{3}{c}{Reinhard~\etal~\cite{reinhard}} && \multicolumn{3}{c}{Photomatix~\cite{Photomatix}} \\ \cmidrule{2-4} \cmidrule{6-8}
        & PSNR $\left( \uparrow \right)$ & SSIM $\left( \uparrow \right)$ & LPIPS $\left( \downarrow \right)$ && PSNR $\left( \uparrow \right)$ & SSIM $\left( \uparrow \right)$ & LPIPS $\left( \downarrow \right)$  \\ \midrule
        Lee \etal~\cite{drhdri}     & 27.97             & 0.9584            & 0.1118            && 23.85            & 0.9282            & 0.1362 \\
        Endo \etal~\cite{drtmo}     & 29.99             & 0.9618            & 0.1257            && 25.98            & 0.9451            & 0.1482 \\
        Ours                        & \textbf{32.08}    & \textbf{0.9819}   & \textbf{0.0508}   && \textbf{28.01}   & \textbf{0.9663}   & \textbf{0.0723} \\
        
        \bottomrule
    \end{tabular}
    \caption[Quantitative comparisons on LDR images using Debevec and Malik~\cite{debevec} merging method]{Quantitative comparisons on tone-mapped images with existing methods using Debevec and Malik~\cite{debevec} merging algorithm.}
    \label{tab:compare-ldr-debevec}
\end{table*}

\begin{table*}[!ht]
    \centering
    \small
    \begin{tabular}{lccccccc}
        \toprule
        
        \multirow{2}{*}{Tone-mapping Method} & \multicolumn{3}{c}{Reinhard~\etal~\cite{reinhard}} && \multicolumn{3}{c}{Photomatix~\cite{Photomatix}} \\ \cmidrule{2-4} \cmidrule{6-8}
        & PSNR $\left( \uparrow \right)$ & SSIM $\left( \uparrow \right)$ & LPIPS $\left( \downarrow \right)$ && PSNR $\left( \uparrow \right)$ & SSIM $\left( \uparrow \right)$ & LPIPS $\left( \downarrow \right)$  \\ \midrule
        Lee \etal~\cite{drhdri}     & \textbf{25.29}    & 0.9271            & 0.1628            && 21.27            & 0.8758            & 0.1986 \\
        Endo \etal~\cite{drtmo}     & 22.49             & 0.9271            & 0.1984            && \textbf{23.49}   & 0.8967            & 0.2350 \\
        Ours                        & 23.89             & \textbf{0.9522}   & \textbf{0.1004}   && 20.34            & \textbf{0.9015}   & \textbf{0.1350} \\
        
        \bottomrule
    \end{tabular}
    \caption[Quantitative comparisons on LDR images using Robertson~\etal~\cite{robertson1999merge} merging method]{Quantitative comparisons on tone-mapped images with existing methods using Robertson~\etal~\cite{robertson1999merge} merging algorithm.}
    \label{tab:compare-ldr-robertson}
\end{table*}

\section{On HDR Merging Methods}
\label{sec:hdr_merging_comparisons} 

We compare different HDR merging methods including the commercial software Photomatix~\cite{Photomatix}, OpenCV's merging based on Debevec and Malik~\cite{debevec} and Robertson~\etal~\cite{robertson1999merge}. The comparison results are shown in \cref{tab:compare-hdr-addition} with the evaluation on HDR images. Additional to results of Photomatix reported in the main paper, we provide evaluations on tone-mapped images when the HDR merging method is Debevec and Malik~\cite{debevec} (\cref{tab:compare-ldr-debevec}) and Robertson~\etal~\cite{robertson1999merge} (\cref{tab:compare-ldr-robertson}).
From these results it can be seen that our method is not sensitive to HDR merging methods as in most cases, our method outperform previous methods. 

Note that in the HDR domain, HDR-VDP-2 is a more preferred metric, which shows that Photomatix has better performance in HDR reconstruction. 
This aligns with our investigation as well, as we found that output images from Photomatix has less visual artifacts than those from Debevec and Malik~\cite{debevec} and Robertson~\etal~\cite{robertson1999merge}.

\begin{table}
    \centering
    \small
    \begin{tabular}{cllll}
        \toprule
        \multicolumn{2}{c}{}               & Ours & Endo~\cite{drtmo} & Lee~\cite{drhdri}        \\ \midrule
        \multirow{3}{*}{$\text{EV}_{-2.0}$} & PSNR  & \textbf{27.25}      & 24.40       & 19.40  \\ 
                                            & SSIM  & \textbf{0.9350}     & 0.9022      & 0.7848 \\ 
                                            & LPIPS & \textbf{0.1032}     & 0.1966      & 0.2214 \\ \midrule
        \multirow{3}{*}{$\text{EV}_{-1.5}$} & PSNR  & \textbf{28.13}      & 25.01       & \NA    \\
                                            & SSIM  & \textbf{0.9511}     & 0.9205      & \NA    \\
                                            & LPIPS & \textbf{0.0898}     & 0.1877      & \NA    \\ \midrule
        \multirow{3}{*}{$\text{EV}_{-1.0}$} & PSNR  & \textbf{30.12}      & 26.65       & 23.38  \\
                                            & SSIM  & \textbf{0.9667}     & 0.9386      & 0.9077 \\
                                            & LPIPS & \textbf{0.0732}     & 0.1764      & 0.1323 \\ \midrule
        \multirow{3}{*}{$\text{EV}_{-0.5}$} & PSNR  & \textbf{33.92}      & 30.23       & \NA    \\
                                            & SSIM  & \textbf{0.9823}     & 0.9569      & \NA    \\
                                            & LPIPS & \textbf{0.0509}     & 0.1602      & \NA    \\ \midrule
        \multirow{3}{*}{$\text{EV}_{+0.5}$} & PSNR  & \textbf{33.50}      & 32.65       & \NA    \\
                                            & SSIM  & 0.9629     & \textbf{0.9645}      & \NA    \\
                                            & LPIPS & \textbf{0.0379}     & 0.1286      & \NA    \\ \midrule
        \multirow{3}{*}{$\text{EV}_{+1.0}$} & PSNR  & \textbf{31.15}      & 30.61       & 25.28  \\
                                            & SSIM  & \textbf{0.9613}     & 0.9552      & 0.9290 \\
                                            & LPIPS & \textbf{0.0553}     & 0.1335      & 0.1207 \\ \midrule
        \multirow{3}{*}{$\text{EV}_{+1.5}$} & PSNR  & \textbf{29.74}      & 29.19       & \NA    \\
                                            & SSIM  & \textbf{0.9541}     & 0.9442      & \NA    \\
                                            & LPIPS & \textbf{0.0734}     & 0.1405      & \NA    \\ \midrule
        \multirow{3}{*}{$\text{EV}_{+2.0}$} & PSNR  & \textbf{29.04}      & 28.12       & 21.84  \\
                                            & SSIM  & \textbf{0.9468}     & 0.9324      & 0.8840 \\
                                            & LPIPS & \textbf{0.0899}     & 0.1469      & 0.1940 \\
        \bottomrule
    \end{tabular}
    \caption[Quantitative results on inferred bracketed images]{
    Quantitative results on inferred bracketed images using $\text{EV}_0$ as the input. 
    The {\NA} symbol indicates \emph{Not Available} due to the proposed model by Lee \etal~\cite{drhdri} can only produce images with $\text{EV}$ differ by one value.
    }
    \label{tab:bracket-comparison}
\end{table}

\section{Additional Results and Comparisons}
\label{sec:more_comparisons} 
\noindent\textbf{Quantitative comparisons on multi-exposure images.}
We use $\text{EV}_0$ as input and compare the predicted bracketed images as shown in \cref{tab:bracket-comparison}. 
As can be seen, our model's scores are better than other methods. This result can be interpreted as that the overall structure of images using our model is well reconstructed as well as perceptually similar to ground-truths. 

\begin{figure}
    \centering

    \begin{subfigure}{\linewidth}
        \centering
        \includegraphics[width=\linewidth]{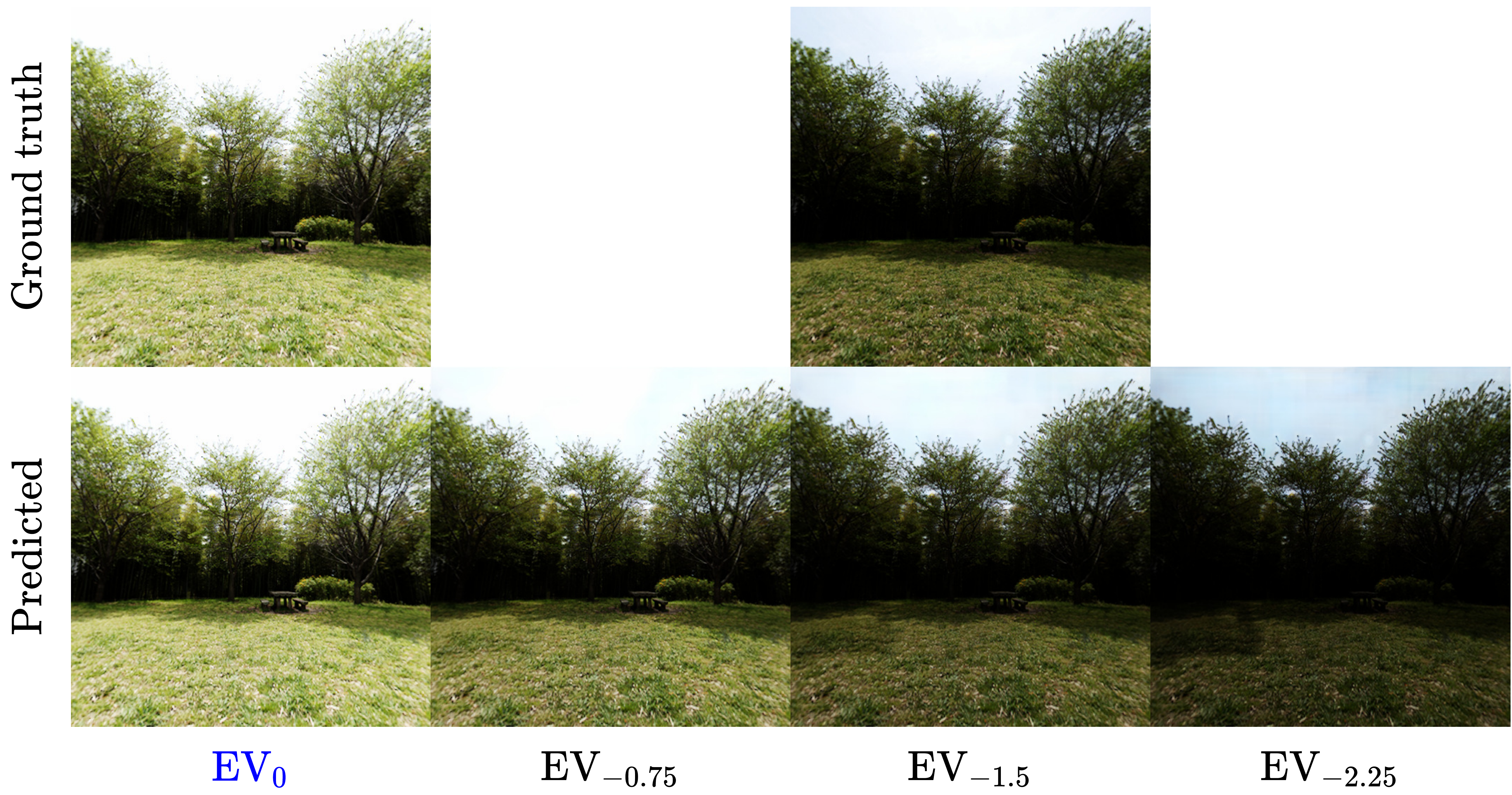}
        \caption[Down-exposed images inferred using our framework]{Down-exposed inferred images.}
        \label{fig:example-inferred-down-arbitrary}
    \end{subfigure}
    
    \begin{subfigure}{\linewidth}
        \centering
        \includegraphics[width=\linewidth]{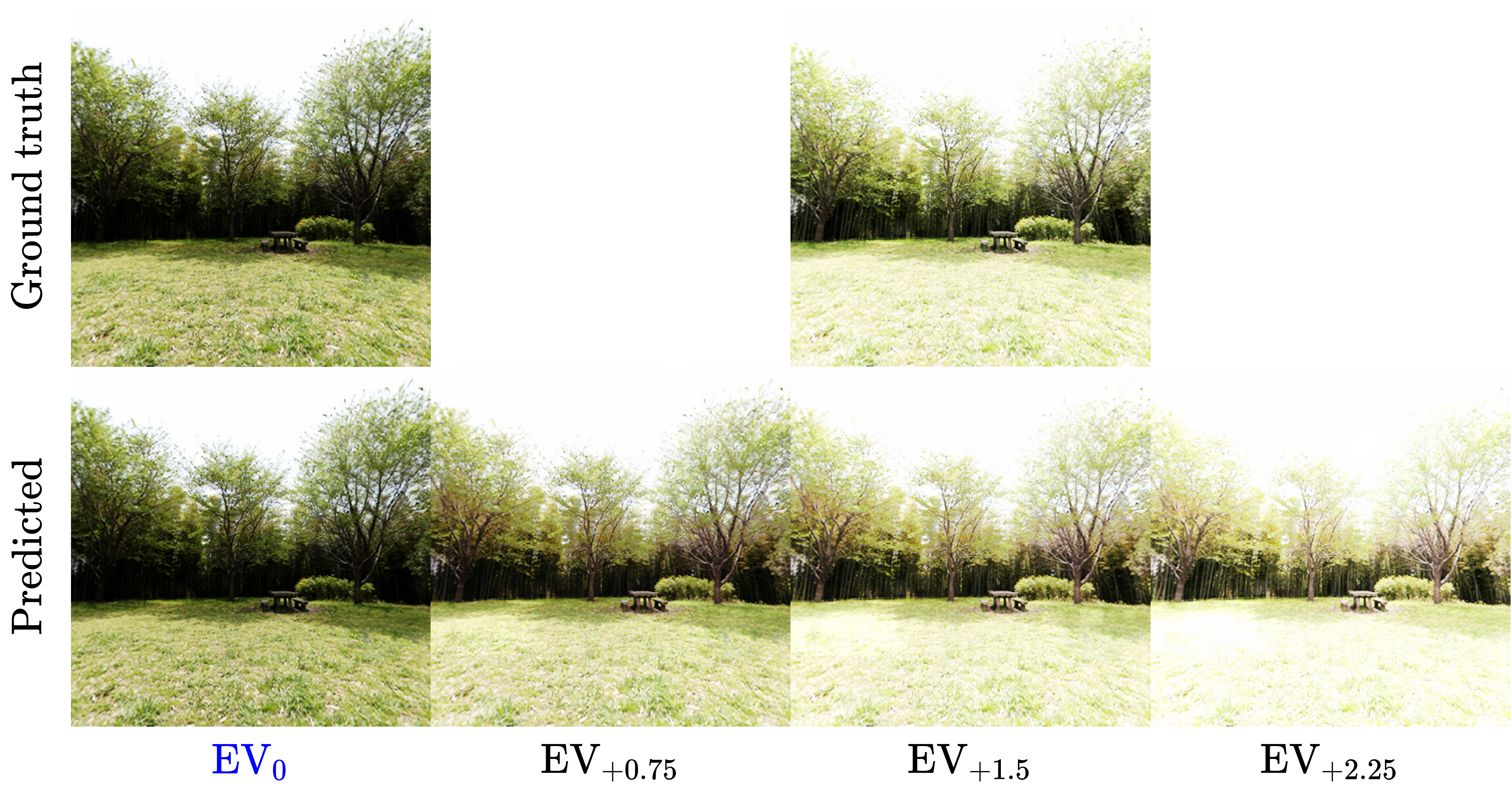}
        \caption[Up-exposed images inferred using our framework]{Up-exposed inferred images.}
        \label{fig:example-inferred-up-arbitrary}
    \end{subfigure}
    
    \caption{Our method can generate virtual exposures at arbitrary exposure value not in the training set, e.g., $\text{EV}_{\pm 0.75}$ and $\text{EV}_{\pm 2.25}$. \textcolor{blue}{Blue} text indicates the input image ($\text{EV}_{0}$).}
    \label{fig:example-inferred-arbitrary}
\end{figure}

\noindent\textbf{Qualitative comparisons on multi-exposure images.}
A significant advantage of our method compared to the previous works is that our approach allows the synthesis of images at arbitrary exposure values. 
We demonstrate this capability in \cref{fig:example-inferred-arbitrary}, where the exposure value can be 0.75, 1.5, and 2.25. 
In the included video, our network can predict a smooth change in the virtual exposures of a scene. 

\begin{figure*}
    \centering
    \includegraphics[width=\linewidth]{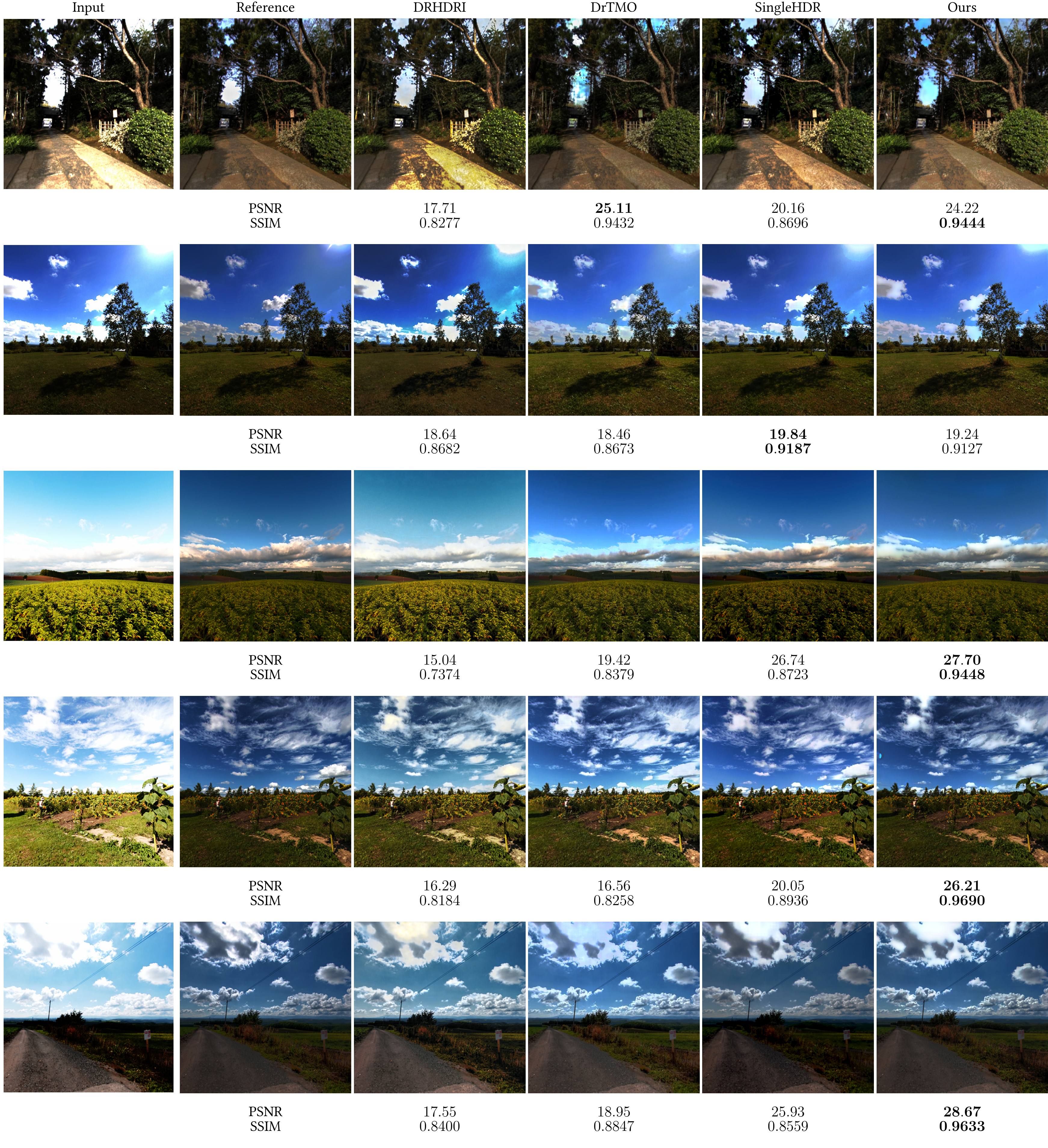}
    \caption{Tone-mapped HDR images comparison between ours, DrTMO~\cite{drtmo}, Deep Recursive HDRI~\cite{drhdri}, and SingleHDR~\cite{singlehdr} on nature scenes.}
    \label{fig:compare-tone-mapped-supp-nature}
\end{figure*}

\begin{figure*}
    \centering
    \includegraphics[width=\linewidth]{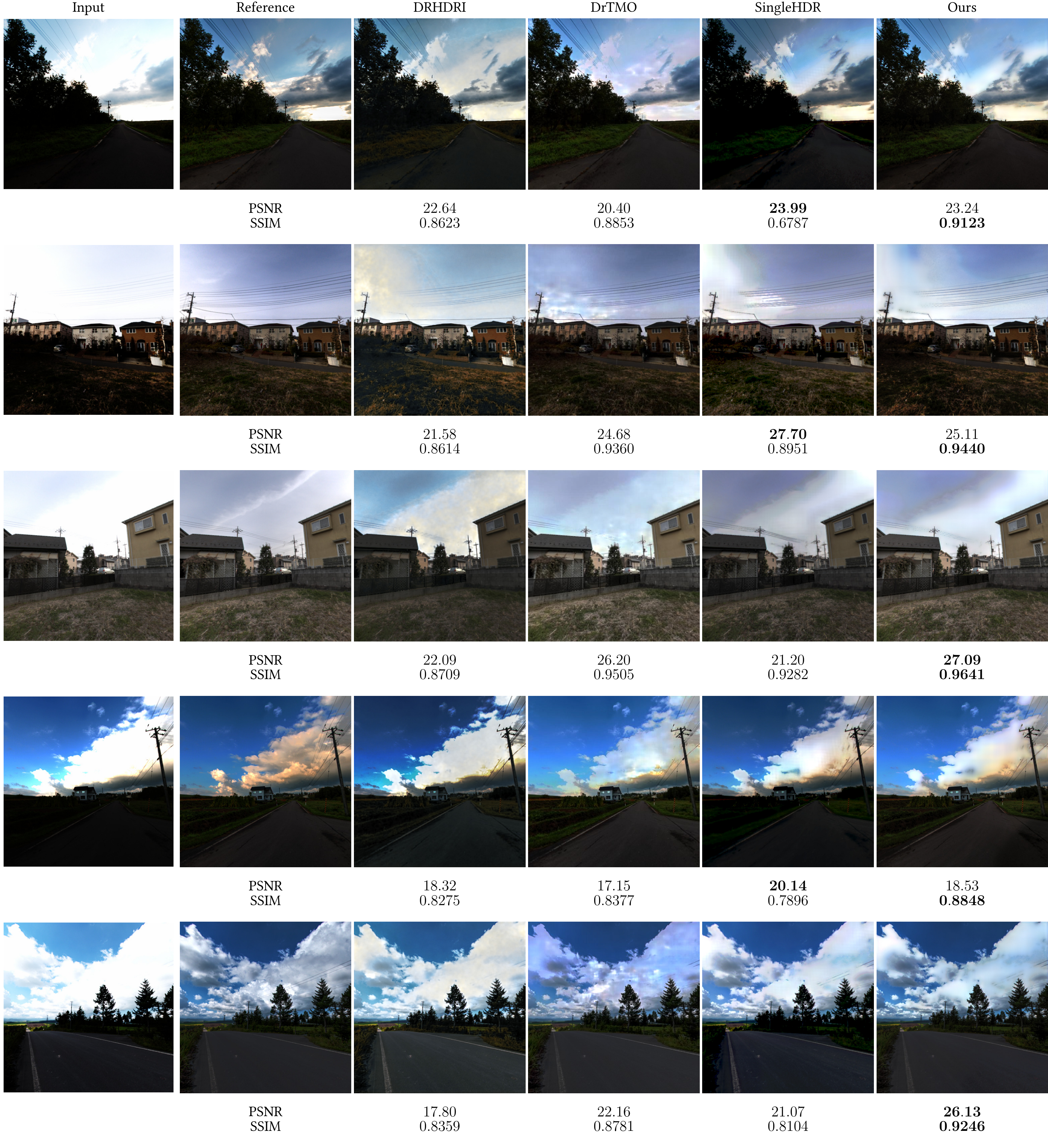}
    \caption{Tone-mapped HDR images comparison between ours, DrTMO~\cite{drtmo}, Deep Recursive HDRI~\cite{drhdri}, and SingleHDR~\cite{singlehdr} on outdoor scenes.}
    \label{fig:compare-tone-mapped-supp-sky}
\end{figure*}

\begin{figure*}
    \centering
    \includegraphics[width=\linewidth]{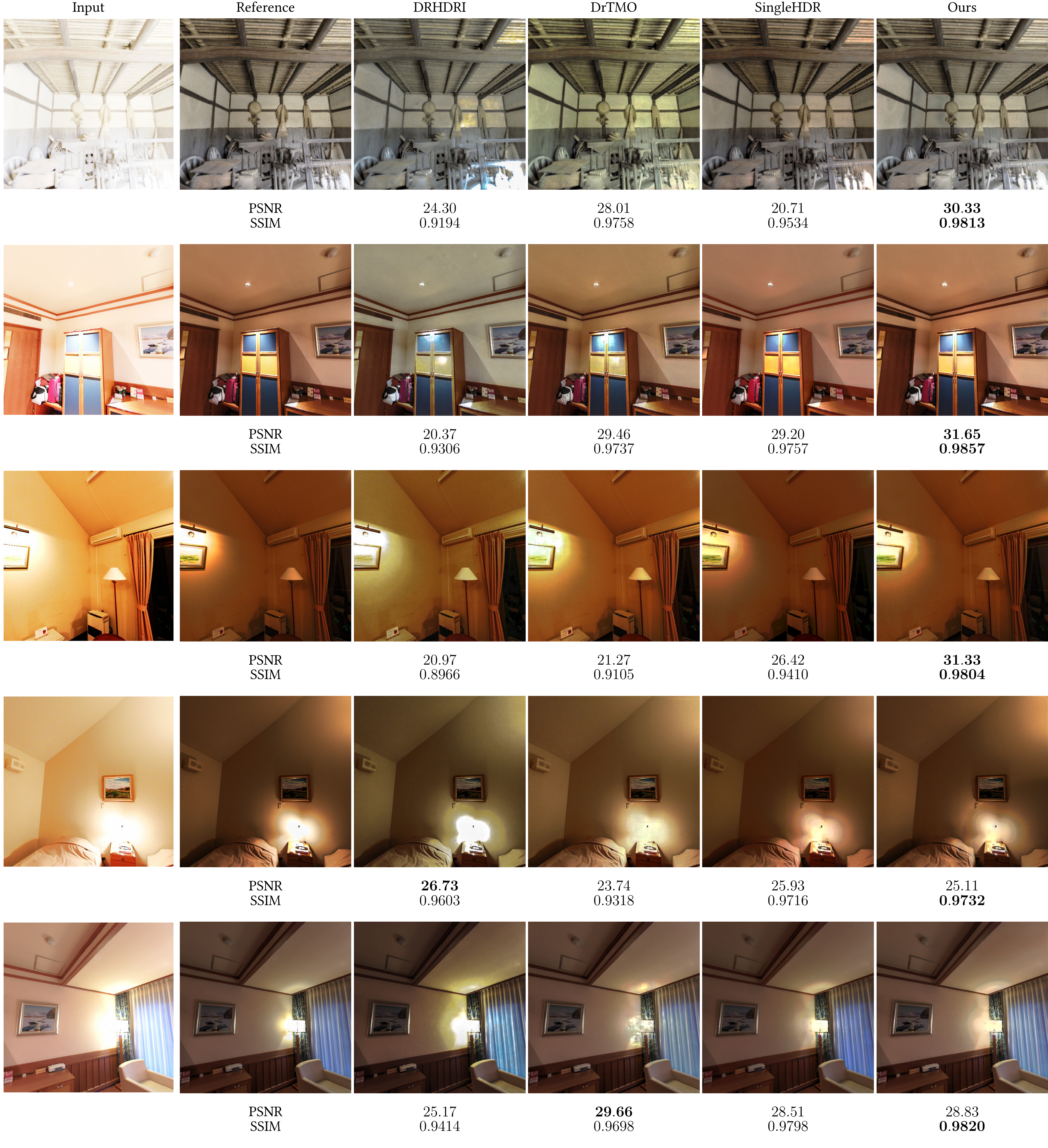}
    \caption{Tone-mapped HDR images comparison between ours, DrTMO~\cite{drtmo}, Deep Recursive HDRI~\cite{drhdri}, and SingleHDR~\cite{singlehdr} on indoor scenes.}
    \label{fig:compare-tone-mapped-supp-indoor}
\end{figure*}

\noindent\textbf{Qualitative comparisons on tone-mapped images.}
We give further qualitative evaluation of tone-mapped images on diverse scenes as shown in \cref{fig:compare-tone-mapped-supp-nature} (natural scenes), \cref{fig:compare-tone-mapped-supp-sky} (outdoor scenes), and \cref{fig:compare-tone-mapped-supp-indoor} (indoor scenes). Our method can reconstruct the HDR that matches well to the reference in color and contrast as well as produce minimal artifacts compared to DrTMO~\cite{drtmo}, DRHDRI~\cite{drhdri}, and SingleHDR~\cite{singlehdr} method.

\begin{figure*}
    \centering
    
    \begin{subfigure}{\textwidth}
        \centering
        \includegraphics[width=\linewidth]{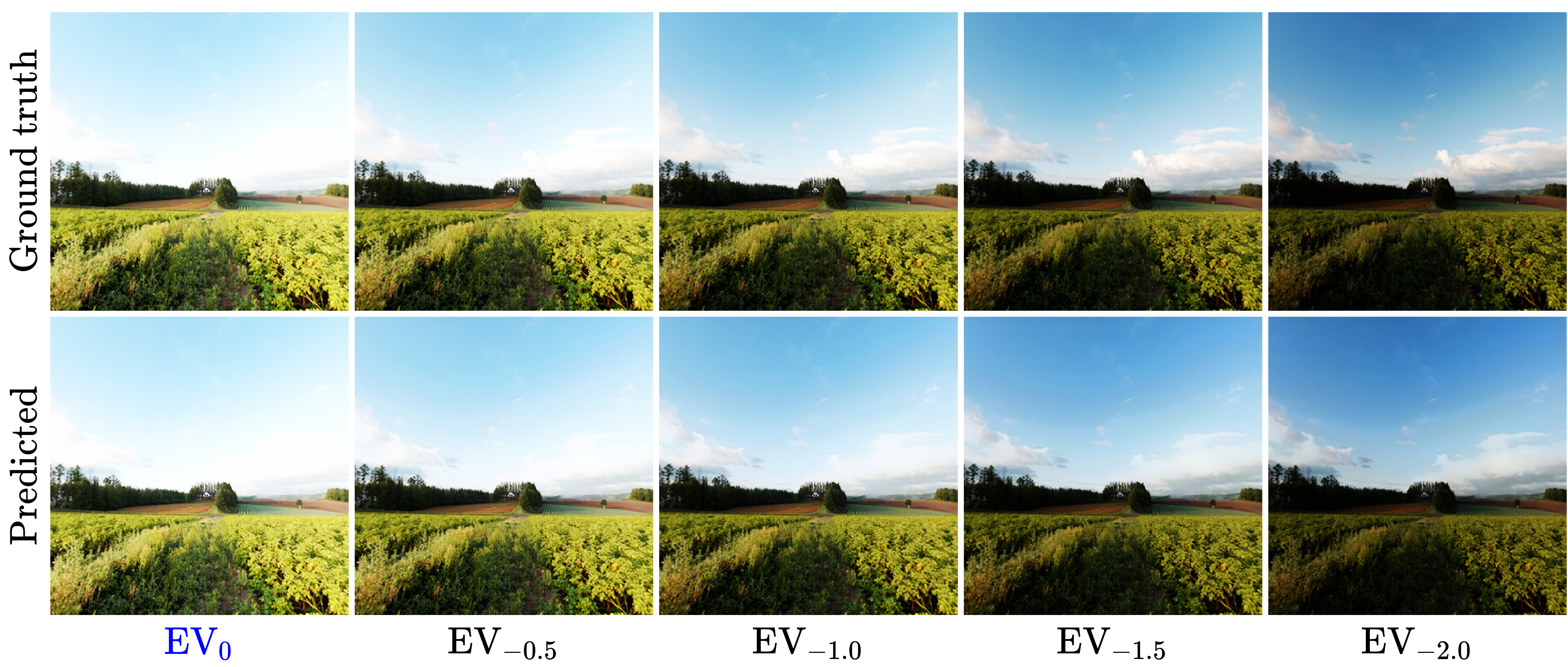}
        \caption{Down-exposure inferred images.}
        \label{fig:example-inferred-down}
    \end{subfigure}
    
    \begin{subfigure}{\textwidth}
        \centering
        \includegraphics[width=\linewidth]{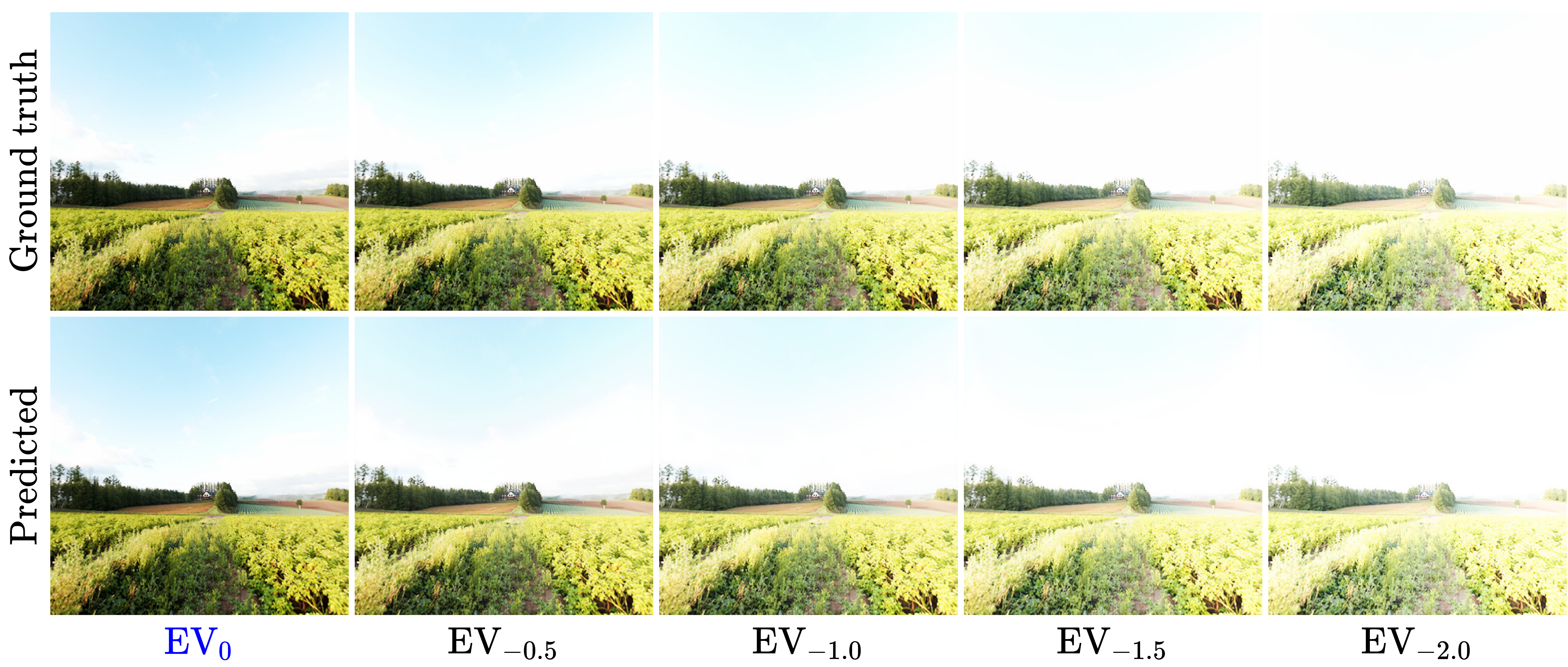}
        \caption{Up-exposure inferred images.}
        \label{fig:example-inferred-up}
    \end{subfigure}
    
    \caption{
        Predicted multi-exposure images. 
        The predicted stack matches the ground truth very well in terms of color and contrast. 
        Input is \blue{$\text{EV}_{0}$}.
    }
    \label{fig:example-inferred}
\end{figure*}

\noindent\textbf{Application: virtual bracketed images.} 
Our method can be used as an application for virtually changing the exposure for an input image. 
\Cref{fig:example-inferred} illustrates the generated bracketed images by ours along with the references. 
We can see that the predicted stack matches the ground truth very well in terms of color and contrast.
Our model can also generate images with smooth changes in the exposure values that are not defined in the training dataset. Please refer to supplementary materials for a video demonstration.

\section{Details of the Proposed Network}
\label{sec:network_details}

\subsection{Network Architecture}
\label{sec:network_architecture}
As our goal is to generate different exposure images from the input image, this can be seen as an image-to-image translation task. 
We adopt the U-Net~\cite{unet} like architecture with the encoder-decoder module, which has shown good performance in this task. 
When the data goes to the next level, the size of feature maps is reduced by half, vertically and horizontally, and conversely doubled. 
Then, the abstracted feature map is reassembled with the previous feature maps for creating the desired output through a structure that increases the width and height of the feature map. 
In this structure, we add skip-connections between encoder and decoder layers so that the characteristics of low-level features are reflected in the output. 
The down-sampling block consists of a convolutional layer followed by one batch normalization layer~\cite{batchnorm} and $\text{ReLU}$ activation function. 
The up-sampling block contains a sub-pixel convolution layer~\cite{subpixelconv} and one convolutional layer instead of a deconvolution layer or resized convolution~\cite{odena2016deconvolution} as Aitken \etal~\cite{subpixelconv} showed that the sub-pixel convolution has more modeling power with the same computation as resize convolution~\cite{odena2016deconvolution}.
The convolutional layer's output then gets passed onto the batch normalization layer and one $\text{Leaky ReLU}$ activation function. 
The same architecture is used for all three sub-networks except for the output convolutional layer of these.

In our implementation, our choice of U-Net for each sub-network consists of 7 levels. 
Each level has two convolution layers with a kernel size of $3\times3$, a stride of 1, and padding of 1. 
The input is first extracted into 16 and 32 features in HDR Encoding Net and Up/Down-Exposure Net, respectively. 
The number of features then doubled at each level until reaching 256 features for HDR Encoding Net and 512 for Up/Down-Exposure Net. 
In the decoder, the extracted features get channel-wise concatenation from the previous level encoder at each level's start. 
The last convolution layer in each sub-network applies $1\times1$ kernel to combine feature maps. We define each sub-network to produce a 3-channel output.

\begin{figure*}[!ht]
    \centering
    \includegraphics[width=\linewidth]{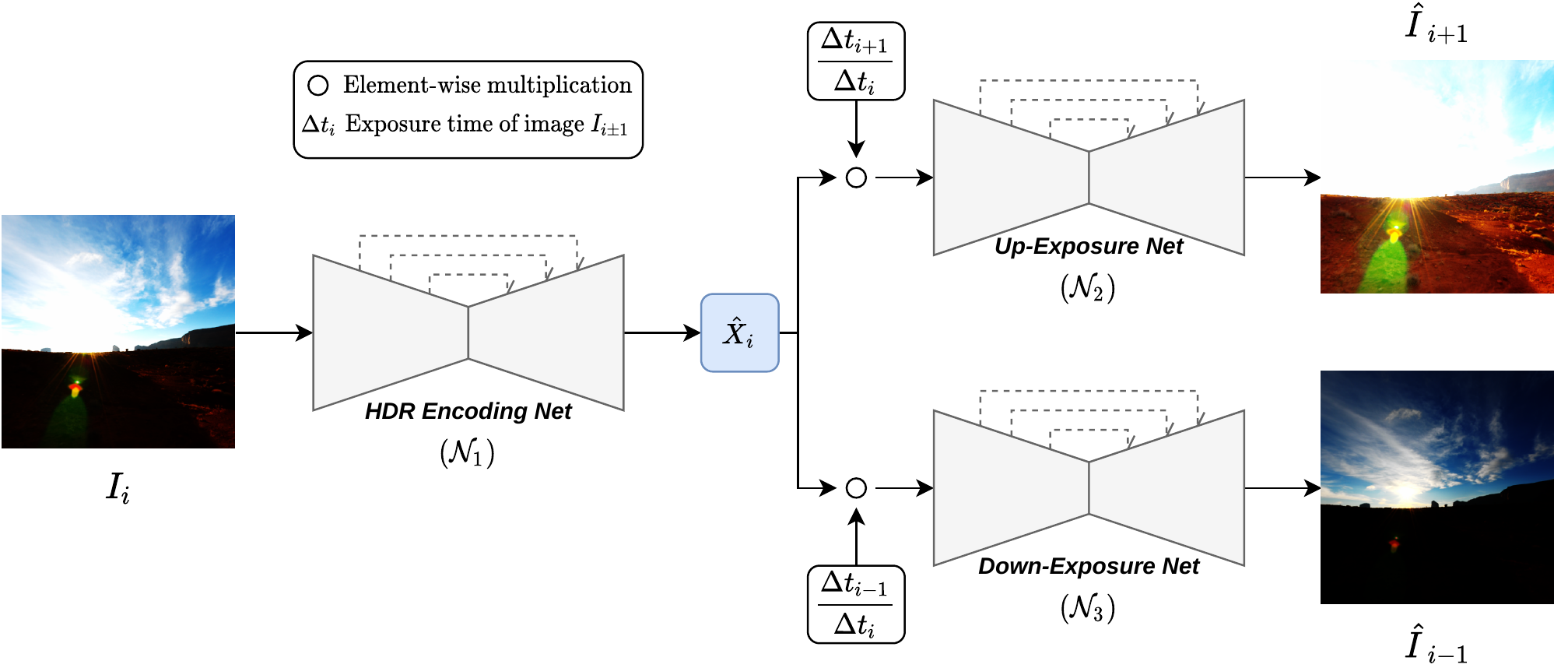}
    \caption{At inference, we take a single image as input and predict multiple exposures by passing the image to Up-Exposure and Down-Exposure Net and varying the exposure time $\Delta t$. The HDR image can be reconstructed from multiple exposures and tone-mapped using traditional methods.}
    \label{fig:inference-process}
\end{figure*}

\noindent\textbf{HDR Encoding Net.}
In the \emph{HDR Encoding Net}, the last layer contains one convolutional layer followed by $\tanh$ activation function and normalization. 
Mathematically, given the output features from the convolutional layer as $F$, our network $\mathcal{N}_1$ output is:
\begin{align}
    \hat{X} = \frac{1}{3}\left( \tanh(F) + I + 1 \right)
\end{align}
with $I$ is the input image for this network, and $\hat{X}$ is the sensor exposure representation of its image. 
The $\tanh$ activation function's output gets added with the input image before feeding into the following network. 
As $\tanh$'s output value is within the range of $\left[ -1, 1 \right]$ thus can be seen as a global adjustment on the input image $I$. 
With this, the network can better find a representation that is suitable to generate different exposure images. 
Then, as the physical property of sensor irradiance can only have positive values, we normalize the output to scale negative values after the previous adjustment.
We experimented and decide to use the $\tanh$ activation function instead of $\text{ReLU}$ as it showed stable training and faster convergence under the use of $\tanh$.

\noindent\textbf{Up/Down-Exposure Net.}
As for Up and Down-Exposure Net, the two sub-networks have to output the longer and shorter exposures, respectively. For this reason, the last layer is the combination between a convolutional layer followed by the \emph{normalized $\tanh$} activation function that is defined as:
\begin{align}
    \tanh_{\text{norm}}(x) = \frac{1}{2} \left( \tanh{(x)} + 1 \right)
\end{align}
The activation output of the function will lie within $\left[ 0, 1 \right]$, which can be seen as the normalized image. 
We opt for this function as $\tanh$ has stronger gradients than the usual sigmoid $\sigma{(x)}$ which helps speed up the learning process.
We then can use other reconstruction loss functions to optimize the network.

\begin{figure}
    \centering
    \includegraphics[width=\linewidth]{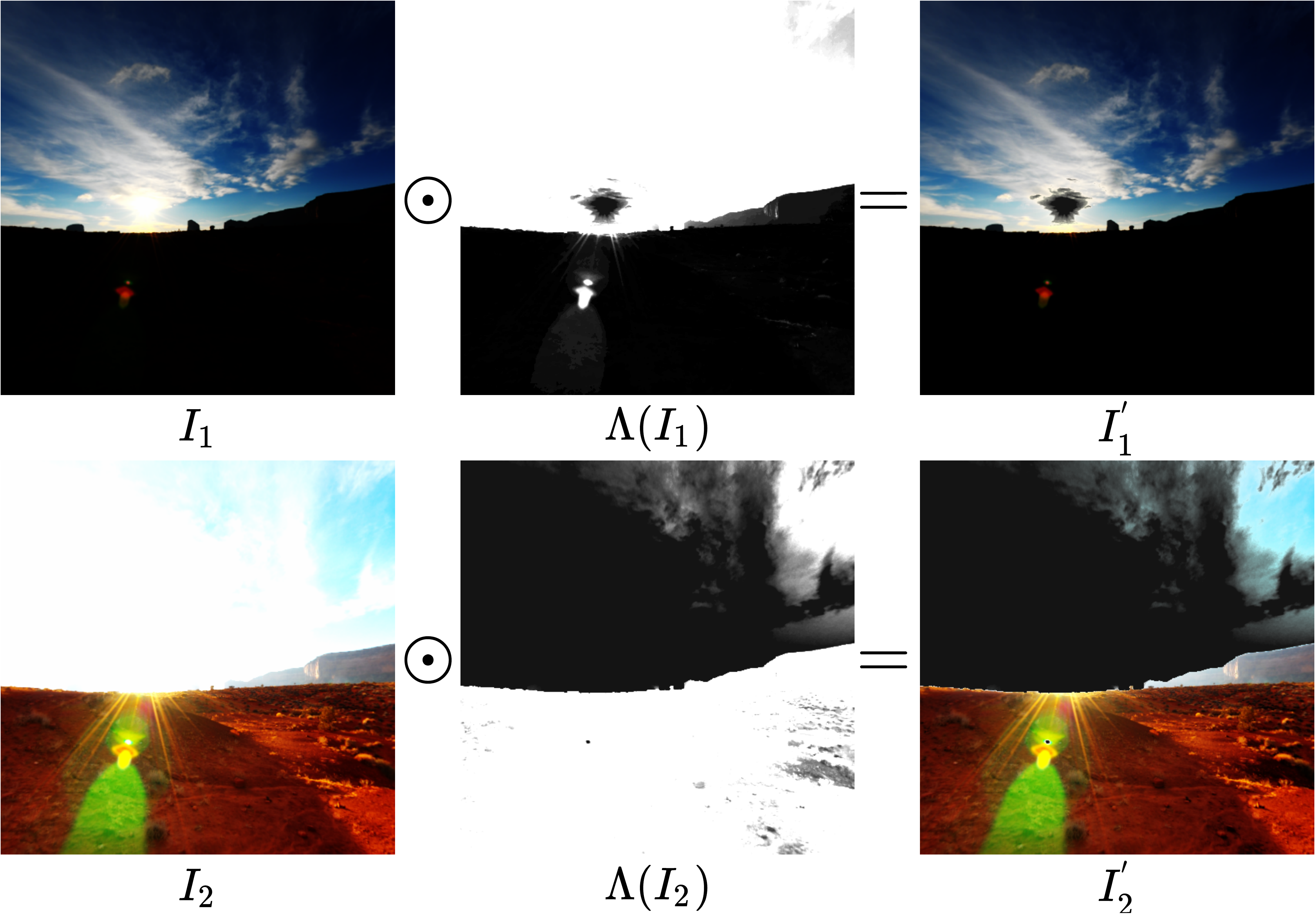}
    \caption{An example of the masks that are applied to the input images before passing through our model.}
    \label{fig:mask-illustration}
\end{figure}

\noindent\textbf{Masked Regions.}
As mentioned before, our model takes input with masked over- and under-exposed regions.
The input images first convert from RGB into YUV color space.
Then based on the luma component $Y$ of the image, we identify which pixel is over- or under-exposed.
Particularly, we denote $\Lambda\left( . \right)$ as the indicator function for the well-exposed regions of image:
\begin{align}
    I^{'}_{i} &= I_{i} \odot \Lambda\left( I_i \right), ~i = 1, 2
\end{align}
The function outputs a soft mask within the range $\left[ 0, 1 \right]$ that helps to define how well-exposed each pixel is. The value of 1 indicates that the pixel is well-exposed. Conversely, the 0 value is assigned to the pixels that are under-exposed or saturated. Mathematically, $\Lambda\left( I \right)$ is formulated as:
\begin{align}
    \Lambda_1\left( I \right) &= 1 - \displaystyle{\frac{\max{\left( 0, \left( 1 - \gamma \right) - I \right)}}{1 - \gamma}} \nonumber\\
    \Lambda_2\left( I \right) &= 1 - \displaystyle{\frac{\max{\left( 0, I - \gamma \right)}}{1 - \gamma}} \nonumber\\
    \Lambda\left( I \right) &= \max{\left( \Lambda_1\left( I \right), \Lambda_2\left( I \right)\right)}
\end{align}
where $\gamma$ is the threshold to determine whether a pixel is over/under-exposed or not. 
In our implementation, we choose $\gamma = 0.05$.
An example of the masks is shown in \cref{fig:mask-illustration}.

\subsection{Inference process}
\label{sec:inference-process}
At inference, we take the input image and pass it to the network to generate multiple exposures. Our network allows us to vary the exposure time to obtain an image bracket with exposures from $\text{EV}_{-2}$ to $\text{EV}_{+2}$, respectively. We then apply traditional HDR reconstruction to merge the predicted exposures into the final HDR image. 
\Cref{fig:inference-process} demonstrates our inference process. 

\end{document}